\documentclass[aps,a4paper,onecolumn,amsmath,amsfonts,preprintnumbers,nofootinbib,superscriptaddress,eqsecnum,secnumarabic]{revtex4-2}
\usepackage[a4paper, hdivide={1.91cm,,1.165cm}, vdivide={1.83cm,,3.8cm}]{geometry}

\usepackage{caption}
\usepackage{ragged2e}
\DeclareCaptionJustification{justified}{\justifying}
\usepackage{amssymb,amsmath,hhline,graphicx}
\usepackage[bookmarks=true,bookmarksnumbered=true,colorlinks=true,urlcolor=black, citecolor=black,linkcolor=black,breaklinks=true]{hyperref}
\usepackage{ulem}
\usepackage{youngtab}
\usepackage[english]{babel}
\usepackage[T1]{fontenc}
\makeatletter\AtBeginDocument{\let\@elt\relax}\makeatother
\usepackage{float}
\usepackage{braket}
\usepackage{amsfonts,bbm}
\usepackage{mathrsfs}
\usepackage{dsfont,tensor}
\usepackage{enumerate}
\usepackage{color}
\usepackage{xcolor}
\usepackage{array}
\usepackage{booktabs}
\usepackage{units}
\usepackage{textcomp}
\usepackage{mathtools}
\usepackage{cancel}
\usepackage{slashed}
\usepackage{multirow}
\usepackage{comment}
\usepackage{relsize}
\usepackage{bm}
\usepackage{bbold}
\usepackage{subfig}

\renewcommand{\i}{\text{i}}
\newcommand{\dd}{\text{d}}

\newcommand{\mx}{m_{\chi}}
\newcommand{\tx}{T_{\chi}}

\newcommand{\pts}[1]{\phantom{.}\hfill(\textit{#1~point}\ifthenelse{\equal{#1}{1}}{}{\textit{s}})}

\newcommand{\Msun}{{\ifmmode{{\rm{M_{\odot}}}}\else{${\rm{M_{\odot}}}$}\fi}}

\newcommand{\beq}{\begin{equation}}
\newcommand{\eeq}{\end{equation}}
\newcommand{\bea}{\begin{eqnarray}}
\newcommand{\ena}{\end{eqnarray}}

\newcommand{\lsim}{\mathrel{\mathop{\kern 0pt \rlap
{\raise.2ex\hbox{$<$}}}
\lower.9ex\hbox{\kern-.190em $\sim$}}}
\newcommand{\gsim}{\mathrel{\mathop{\kern 0pt \rlap
{\raise.2ex\hbox{$>$}}}
\lower.9ex\hbox{\kern-.190em $\sim$}}}
\begin{document}

\title{Evaporation of dark matter from celestial bodies} 
\author{Raghuveer Garani}
\email{garani@fi.infn.it}
\affiliation{INFN Sezione di Firenze, Via G. Sansone 1, I-50019 Sesto Fiorentino, Italy}

\author{Sergio Palomares-Ruiz}
\email{sergiopr@ific.uv.es}
\affiliation{Instituto de F\'{\i}sica Corpuscular (IFIC), Universitat de Val\`encia -- CSIC, Parc Científic, C/ Catedrático José Beltrán, 2, E-46980 Paterna, Spain}

\begin{abstract}	
\vspace{2mm}	
Scatterings of galactic dark matter (DM) particles with the constituents of celestial bodies could result in their accumulation within these objects. Nevertheless, the finite temperature of the medium sets a minimum mass, the evaporation mass, that DM particles must have in order to remain trapped. DM particles below this mass are very likely to scatter to speeds higher than the escape velocity, so they would be kicked out of the capturing object and escape. Here, we compute the DM evaporation mass for all spherical celestial bodies in hydrostatic equilibrium, spanning the mass range $[10^{-10} - 10^2]~M_\odot$, for constant scattering cross sections and $s$--wave annihilations. We illustrate the critical importance of the exponential tail of the evaporation rate, which has not always been appreciated in recent literature, and obtain a robust result: for the geometric value of the scattering cross section and for interactions with nucleons, at the local galactic position, the DM evaporation mass for all spherical celestial bodies in hydrostatic equilibrium is approximately given by $E_c/\tx \sim 30$, where $E_c$ is the escape energy of DM particles at the core of the object and $\tx$ is their temperature. In that case, the minimum value of the DM evaporation mass is obtained for super-Jupiters and brown dwarfs, $m_{\rm evap} \simeq 0.7$~GeV. For other values of the scattering cross section, the DM evaporation mass only varies by a factor smaller than three within the range $10^{-41}~\textrm{cm}^2 \leq \sigma_p \leq 10^{-31}~\textrm{cm}^2$, where $\sigma_p$ is the spin-independent DM-nucleon scattering cross section. Its dependence on parameters such as the galactic DM density and velocity, or the scattering and annihilation cross sections is only logarithmic, and details on the density and temperature profiles of celestial bodies have also a small impact.
\end{abstract}

\maketitle

\section{Introduction}
\label{sec:introduction}

The possibility of capture of dark matter (DM) particles by celestial bodies has a long history dating back to the mid 1980s~\cite{Press:1985ug, Griest:1986yu, Gould:1987ju, Gould:1987ir}. Already in the late 1970s, some of the potential effects of DM annihilations and scatterings on energy transport within the Sun were realized~\cite{Steigman:1997vs} and then further proposed to alleviate the so-called solar neutrino problem~\cite{Spergel:1984re, Faulkner:1985rm, Krauss:1985ks, Gilliland:1986, Nauenberg:1986em}. Nevertheless, these early works did not consider the process of DM capture, but assumed the required amount of accumulated DM particles. The process of capture of galactic DM particles by the Sun and the Earth was first studied in Refs.~\cite{Press:1985ug, Griest:1986yu, Gould:1987ju, Gould:1987ir}, which set the ground of further calculations. 

The scattering of DM particles in galactic halos with the nuclei (or electrons) of celestial objects could bring those particles into close orbits and finally result in their gravitational capture within the objects. DM particles would undergo further scatterings and, except for very small cross sections, thermalize in a short period of time, so that this process can be approximately considered as instantaneous~\cite{Bertoni:2013bsa, Widmark:2017yvd, Liang:2018cjn, Blennow:2018xwu, Gaidau:2018yws, Garani:2020wge}. In addition to the Sun~\cite{Press:1985ug, Silk:1985ax, Freese:1985qw, Krauss:1985aaa, Hagelin:1986gv, Gaisser:1986ha, Srednicki:1986vj, Griest:1986yu, Gould:1987ju}, capture of DM particles and its potential observational consequences have also been considered in other celestial bodies, such as the Earth~\cite{Freese:1985qw, Krauss:1985aaa, Gaisser:1986ha, Gould:1987ir, Gould:1988eq, Gould:1991va}, other solar system planets and satellites~\cite{Krauss:1985aaa, Fukugita:1988uh, Dimopoulos:1989hk, Kawasaki:1991eu, Mitra:2004fh, Adler:2008ky, Bramante:2019fhi, Garani:2019rcb}, exoplanets~\cite{Adler:2008ky, Leane:2020wob}, brown dwarfs~\cite{Zentner:2011wx, Leane:2020wob}, main-sequence~\cite{Faulkner:1985rm, Finzi:1986bh, Faulkner:1988, Salati:1989, Bouquet:1989, DeLuca:1989eh, Fairbairn:2007bn, Iocco:2008xb, Freese:2008ur, Iocco:2008rb, Freese:2008wh, Yoon:2008km, Taoso:2008kw, Scott:2008ns} and post-main sequence~\cite{Renzini:1987, Finzi:1987, Spergel:1988, Dearborn:1990} stars, and compact objects such as white dwarfs~\cite{Moskalenko:2006mk, Moskalenko:2007ak, Bertone:2007ae, McCullough:2010ai, Bramante:2015cua, Graham:2015apa} and neutron stars~\cite{Goldman:1989nd, Gould:1990b, Kouvaris:2007ay, Bertone:2007ae, Kouvaris:2010vv, deLavallaz:2010wp, Baryakhtar:2017dbj}. 

Below a minimum DM mass, scatterings of thermalized DM particles off ambient targets (nuclei or electrons) would boost them very efficiently to speeds above the escape velocity, so that they would evaporate from the capturing celestial body. Therefore, only for DM masses above the evaporation mass, effects of the capture of DM particles could have any impact on the evolution of the objects or could produce any other observable signature (the exception being the potential signal of evaporated DM particles at low-threshold direct detection experiments~\cite{Kouvaris:2015nsa}). The DM evaporation mass has been computed at different levels of detail for the case of the Sun~\cite{Steigman:1997vs, Spergel:1984re, Faulkner:1985rm, Krauss:1985aaa, Griest:1986yu, Gaisser:1986ha, Gould:1987ju, Nauenberg:1986em, Busoni:2013kaa, Liang:2016yjf, Garani:2017jcj, Busoni:2017mhe}, but has been less studied for other celestial bodies. Some examples, however, are the calculations for the Earth~\cite{Freese:1985qw, Krauss:1985aaa, Gould:1988eq, Garani:2019rcb}, the Moon~\cite{Garani:2019rcb},  Mars~\cite{Bramante:2019fhi}, giant planets~\cite{Leane:2020wob, Leane:2021tjj}, brown dwarfs~\cite{Zentner:2011wx, Leane:2020wob}, main-sequence stars~\cite{Zentner:2011wx, Raen:2020qvn}, horizontal-branch stars~\cite{Spergel:1988, Gould:1990}, and neutron stars~\cite{Garani:2018kkd, Bell:2020lmm}.

The process of DM evaporation from celestial bodies depends on the temperature of potential scatterers in the medium, which are assumed to be in thermal equilibrium. Part of the finite thermal energy of nuclei (or electrons) would be transferred to DM particles via elastic scatterings, which could then end up with energies above the local escape energy. Unless further scatterings take place, these DM particles would not be gravitationally bound any more and thus would escape from the object. Obviously, in the limit of zero temperature, the evaporation rate is zero. 

Light thermalized DM particles typically have speeds higher than heavier particles, so they are more likely to end up with speeds higher than the local escape velocity after scattering off target particles in the medium. The combination of the DM velocity distribution along with the probability to escape after scattering results in the evaporation rate having an exponential dependence~\cite{Griest:1986yu, Gould:1987ju, Gould:1990}. The key quantity is the exponent. In the thin regime (long mean free path), evaporation mostly occurs close to the center of the object,\footnote{In the thick regime (short mean free path), DM particles approach local thermodynamic equilibrium and evaporation occurs mostly in a shell closer to the surface of the object~\cite{Gould:1990}, where the temperature, but also the escape velocity, are lower. In any case, qualitatively, the discussion is analogous.} and the exponent is $E_c/\tx$, the ratio of the escape energy of DM particles at the center of the object to their temperature (assuming an isothermal distribution). The smaller the exponent, the larger the number of DM particles close to the escape velocity and the higher the probability to gain enough energy to escape, and hence, the higher the evaporation rate. It is a well known fact since the early papers on the topic that, in order to efficiently suppress DM evaporation in the Sun, $E_c/\tx \simeq 30$~\cite{Griest:1986yu}. This implies that the DM evaporation mass is set along the exponential tail of the evaporation rate~\cite{Spergel:1984re, Gaisser:1986ha, Griest:1986yu, Gould:1987ju}. For the case of the Sun, $m_{\rm evap} \simeq 3$~GeV~\cite{Busoni:2013kaa, Garani:2017jcj, Busoni:2017mhe}, whereas for the Earth, $m_{\rm evap} \simeq 12$~GeV~\cite{Freese:1985qw, Krauss:1985aaa, Gould:1988eq, Garani:2019rcb} and for the Moon, $m_{\rm evap} \simeq 70$~GeV~\cite{Garani:2019rcb}.\footnote{Note that Ref.~\cite{Garani:2019rcb} quotes $m_{\rm evap} \simeq 40$~GeV. Nevertheless, to obtain that DM evaporation mass for the Moon, the core temperature was set to $T_c = 700$~K, instead of $T_c = 1700$~K, as written in the text. For the latter core temperature, $m_{\rm evap} \simeq 70$~GeV. These are the values we use as realistic model references. Thermal models of the interior of the Moon predict a core temperature above 1000~K, but it could reach a value of up to 2000~K~\cite{Garcia:2019}.}

In this work, we compute the DM evaporation mass for a wide range of celestial bodies, from the smallest objects with spherical shape that can be in hydrostatic equilibrium (small satellites and dwarf planets), $M \simeq 10^{-10}~M_\odot$~\cite{Tancredi:2009, Lineweaver:2010gg}, to the most massive main-sequence stars, $M \simeq 100~M_\odot$. In addition, we also discuss the DM evaporation mass for post-main-sequence stars, white dwarfs and neutron stars. We consider a range of DM-nucleon (momentum and velocity independent) scattering cross sections that covers ten decades, $10^{-41}~\textrm{cm}^2 \leq \sigma_p \leq 10^{-31}~\textrm{cm}^2$, and which runs over the thin and thick regimes.

Recently, searches of effects of the accretion of sub-GeV DM particles in giant planets and brown dwarfs have been proposed~\cite{Leane:2020wob, Leane:2021ihh, Leane:2021tjj}. The ideas rely on estimates of the DM evaporation mass for those objects as low as a few MeV, based on their relatively large size and cool temperatures. Likewise, a low DM evaporation mass for the Earth and Mars has also been claimed~\cite{Bramante:2019fhi}. Here, we show that these calculations, which assume constant scattering cross sections, neglect the crucial exponential tail of the evaporation rate and underestimate the DM evaporation mass by at least one order of magnitude, and we argue that the suggested implications for DM masses below the correctly evaluated (properly accounting for the exponential tail) DM evaporation mass do not apply.

This paper is organized as follows. In Section~\ref{sec:basics}, we introduce the basic ingredients required for the calculation of the DM evaporation mass: the capture, annihilation and evaporation rates. Then, we describe in some detail the process of DM evaporation in a generic celestial body and explain the importance of the exponential tails in the determination of the DM evaporation mass. In Section~\ref{sec:celestialbodies}, we describe the main average properties, relevant for the calculation of the DM evaporation mass, of all celestial bodies we consider in this work, spanning twelve orders of magnitude in mass, $10^{-10}~M_\odot \leq M \leq 10^2~M_\odot$. The main results of this work are presented in Section~\ref{sec:results}, where we show the value of the DM evaporation mass for all these objects and describe its dependence on several parameters. Finally, in Section~\ref{sec:conclusions}, we summarize our results and draw our conclusions.

\section{Basics of dark matter evaporation}
\label{sec:basics}

DM particles in a galactic halo could scatter off the material of celestial bodies and lose enough energy such that their velocity ends up smaller than the escape velocity of the object, thus becoming gravitationally captured within. The evolution of the total number of DM particles accumulated in a celestial body, $N_\chi(t)$, is given by
\begin{equation}
\frac{d N_\chi (t)}{dt} = \mathcal{C} - \mathcal{A} \, N^2_\chi(t) - \mathcal{E} \, N_\chi(t) ~, 
\end{equation}
where $\mathcal{C}$, $\mathcal{A}$ and $\mathcal{E}$ are the DM capture, annihilation and evaporation rates, which throughout this work, are taken to be constant in time and are defined in the next subsections. In the above equation, canonical two-body annihilation processes are assumed and DM self-interactions are not considered. The solution reads~\cite{Gaisser:1986ha, Griest:1986yu}
\begin{equation}
\label{eq:NDM}
N_\chi (t) = \mathcal{C} \, \tau_{\rm eq} \, \frac{\tanh(\kappa \, t/\tau_{\rm eq})}{\kappa + \frac{1}{2} \, \mathcal{E} \, \tau_{\rm eq} \, \tanh(\kappa \, t/\tau_{\rm eq})} ~, 
\end{equation}
where $\tau_{\rm eq} = 1/\sqrt{\mathcal{A} \, \mathcal{C}}$ is the equilibration time scale in the absence of evaporation and $\kappa = \sqrt{1 + (\mathcal{E} \, \tau_{\rm eq}/2)^2}$. Equilibrium is reached when $\kappa \, t \gg \tau_{\rm eq}$. Before this occurs, $N_\chi \simeq \mathcal{C} \, t$ and, as discussed below, the DM evaporation mass grows with time. Once equilibrium is attained, if evaporation dominates ($\kappa \gg 1$), $N_\chi \simeq \mathcal{C}/\mathcal{E}$, and the number of accumulated DM particles decreases exponentially with decreasing mass (see below). If evaporation is not efficient, $\kappa \simeq 1$ (yet, $t \gg \tau_{\rm eq}$), the number of DM particles is given by $N_\chi \simeq \mathcal{C} \, \tau_{\rm eq} = \sqrt{\mathcal{C}/\mathcal{A}}$. Bearing this in mind, the DM evaporation mass, $m_{\rm evap}$, can be defined as~\cite{Busoni:2013kaa} (see Refs.~\cite{Steigman:1997vs, Spergel:1984re, Gaisser:1986ha, Gould:1990, Garani:2017jcj} for other definitions)
\begin{equation}
\label{eq:evapmass}
\left|N_\chi (t ; m_{\rm evap}) - \frac{\mathcal{C} (m_{\rm evap})}{\mathcal{E} (m_{\rm evap})}\right| \equiv 0.1\, N_\chi (t ; m_{\rm evap}) ~, 
\end{equation}
which is the mass for which the number of captured DM particles reduces to the solution in the limit of  evaporation dominance, $\mathcal{C}/\mathcal{E}$, at the 10\% level. When equilibrium is reached, this condition becomes time independent and simplifies to $\mathcal{E} (m_{\rm evap}) \, \tau_{\rm eq} (m_{\rm evap}) = 1/\sqrt{0.11}$~\cite{Garani:2017jcj}. The calculation of this minimum mass of DM particles that can get efficiently trapped in celestial bodies depends on the properties of the capturing object, and this is the main goal of this work.

\subsection{Main inputs}

In order to set the ground for the discussion about the evaporation process and the minimum DM mass that can be efficiently captured by different celestial bodies, here we define the three relevant rates introduced above in terms of the DM properties (mass and scattering cross section off nuclei) in the vicinity of the celestial objects (which depend on the local DM density and velocity distribution), the properties of the capturing body (mass and size, density and temperature profiles, and composition) and the kinematics of elastic DM-nuclei scatterings. The general features of the DM capture process by non-degenerate non-relativistic targets are well known in the literature. Here, we only sketch the main ingredients and refer the reader to other works for further details~\cite{Press:1985ug, Griest:1986yu, Gould:1987ju, Gould:1987ir, Gould:1989hm, Gould:1990, Gould:1991va, Bottino:2002pd, Guo:2013ypa, Busoni:2013kaa, Catena:2015uha, Vincent:2015gqa, Liang:2016yjf, Garani:2017jcj, Bramante:2017xlb, Busoni:2017mhe, Dasgupta:2019juq, Ilie:2020vec}.

We consider a celestial body of mass $M$ and radius $R$, composed of non-degenerate non-relativistic elements with mass $m_i$, number density $n_i(r)$ which are assumed to be in local thermodynamic equilibrium described by a Maxwell-Boltzmann distribution of temperature $T(r)$. The total number of targets $i$ is $N_i = (X_i/A_i) \, M$, where $X_i$ and $A_i$ are the mass fraction and the mass of element $i$.

We consider the DM population in the vicinity of this object with mass $\mx$, density $\rho_\chi$ and with velocity distribution $f_{v_{\rm cb}}(u_\chi)$, as seen by an observer moving at speed $v_{\rm cb}$ with respect to the galactic rest frame, 
\begin{equation}
\label{eq:fugen}
f_{v_{\rm cb}}(u_\chi) = \frac{1}{2} \, \int_{-1}^{1} f_{\rm gal}\left(\sqrt{u_\chi^2 + v_{\rm cb}^2 + 2 \, u_\chi \, v_{\rm cb} \, \cos{\theta}}\, \right) \, \dd\cos{\theta} = \sqrt{\frac{3}{2 \pi}} \frac{u_\chi}{v_{\rm cb} \, v_d} \, \left( e^{-\frac{3 \, (u_\chi - v_{\rm cb})^2}{2 \, v_d^2}} - e^{-\frac{3 \, (u_\chi + v_{\rm cb})^2}{2 \, v_d^2}}\right)~,
\end{equation}
where $u_\chi$ is the DM velocity at infinity in the celestial body's rest frame, $\cos{\theta}$ is the angle between the DM and the object velocities and $f_{\rm gal}(u_{\rm gal})$ is the DM velocity distribution in the galactic rest frame, which is usually assumed to be a Maxwell-Boltzmann distribution (the so-called standard halo model) with dispersion velocity $v_d = \sqrt{3/2} \, v_{\rm cb}$ at a given position in the halo. Throughout this paper, we use $v_d = 270~\textrm{km/s}$ and $\rho_{\chi} = 0.4~\textrm{GeV/cm}^3$.

These DM particles may eventually interact with the thermal distribution of the object's targets, with differential scattering cross section $\dd \sigma_i/\dd v$. In this work, we focus on constant (i.e., momentum and velocity independent) total scattering cross sections, which can be written in terms of the DM-proton scattering cross section (or the DM-neutron scattering cross section, that we assume to be equal), for spin-independent (SI) or spin-dependent (SD) interactions, at zero momentum transfer, as 
\begin{eqnarray}
\label{eq:crosssectionsSI}
\sigma_{i}^{\rm SI} & = & \left(\frac{\tilde \mu_{A_i}}{\tilde \mu_p}\right)^2 \,
A_i^2 \, \sigma_{p}^{\rm SI} ~, \\
\label{eq:crosssectionsSD}
\sigma_{i}^{\rm SD} & = & \left(\frac{\tilde \mu_{A_i}}{\tilde \mu_p}\right)^2 \,
\frac{4 \, (J_i +1)}{3 \, J_i} \, \left| \langle S_{p, i} \rangle +
\langle S_{n,i} \rangle \right|^2 \, \sigma_{p}^{\rm SD} ~, 
\end{eqnarray}
where $\tilde{\mu}_{A_i}$ and $\tilde{\mu}_p$ are the reduced masses of the DM-nucleus $i$ and DM-proton systems, $J_i$ is total angular momentum of nucleus $i$, $\sigma_p^{\rm SI, SD}$ is the SI/SD DM-proton scattering cross section, and $\langle S_{p,i}\rangle$ and $\langle S_{n,i}\rangle$ are the expectation values of the spins of protons and neutrons averaged over all nucleons~\cite{Ellis:1987sh, Pacheco:1989jz, Engel:1989ix, Engel:1992bf, Divari:2000dc, Bednyakov:2004xq}. Multiple mediators could exist in some models that couple DM to the visible sector. As long as the total scattering cross section remains (approximately) independent of Mandelstam variables in the relevant momentum transfer regime, the discussion presented below applies.

The differential rate at which a DM particle with velocity $w$ scatters off a target $i$ with velocity $u$ and relative angle $\theta_\chi$, in the laboratory frame, to a final velocity $v$ is given by~\cite{Gould:1987ju}
\begin{equation}
\label{eq:scatteringrate}
\mathcal{R}_i^{\pm} (w \to v) = \int n_i(r) \, \frac{\dd \sigma_i}{\dd v} \, |\boldsymbol{w} - \boldsymbol{u}| \, f_i(\boldsymbol{u},r) \, \dd^3 \boldsymbol{u} = \frac{2}{\sqrt{\pi}} \, \frac{n_i(r)}{u_i^3(r)} \, \int_0^\infty \dd u \, u^2 \, \int_{-1}^1 \dd \cos\theta_\chi \, \frac{\dd \sigma_i}{\dd v} \, |\boldsymbol{w} - \boldsymbol{u}| \, e^{-u^2/u_i^2(r)} ~. 
\end{equation}
The superindex $^\pm$ indicates $v \gtrless w$, $f_i(\boldsymbol{u},r)$ is the velocity distribution of the target particles $i$ (assumed to be Maxwell-Boltzmann), with density $n_i(r)$ and temperature $T(r)$, and $u_i(r) \equiv \sqrt{2 \, T(r)/m_i}$ is the most probable speed of the target particles at position $r$. The relative velocity between the DM and the target particles is given by $|\boldsymbol{w} - \boldsymbol{u}| = \sqrt{w^2 + u^2 - 2 \, w \, u \, \cos{\theta_\chi}}$, where $w$ is related to $u_\chi$ by $w^2(r) = u_\chi^2 + v_e^2(r)$, with $v_e(r)$ the escape velocity at a radial distance $r$ from the center of the object. Analytical expressions for these rates when the nuclear form factor is not included or when the targets temperature is neglected (for the capture process, i.e., for $\mathcal{R}^{-} (w \to v)$) were obtained long ago~\cite{Gould:1987ju, Gould:1987ir}.

\subsection{DM capture by celestial bodies}

With all the above ingredients, the capture rate of DM particles by celestial bodies can be generically written as~\cite{Gould:1987ir, Busoni:2017mhe}
\begin{equation}
\label{eq:captureweak}
\mathcal{C} =\sum_i \int_0^{R} s_{\rm cap}(r) \, 4 \pi \, r^2 \, \dd r \int_0^\infty \dd u_\chi \, \left(\frac{\rho_\chi}{\mx}\right) \, \frac{f_{v_{\rm cb}}(u_\chi)}{u_\chi} \, w(r) \int_0^{v_e(r)} \mathcal{R}_i^- (w \to v) \, \dd v ~, 
\end{equation}
where the sum runs over all possible targets. We have also included the suppression factor $s_{\rm cap}(r)$, which ignores multiple collisions, but allows for a smooth transition between the optically thin (small cross sections) and optically thick regimes (large cross sections)~\cite{Busoni:2017mhe}. Not only $\mathcal{R}^{-}_i(w \to v)$, but the total capture rate per unit volume can also be analytically computed when either the targets temperature is neglected or at finite temperature without including a non-trivial nuclear form factor~\cite{Gould:1987ir}, assuming $s_{\rm cap}(r) = 1$.

Thermal effects have a small impact on capture of DM particles by the nuclei of celestial objects (i.e., $\mx v_e^2/2 \gg T$), although they would be more relevant in the interior of the smallest objects discussed in this work. Moreover, effects from the lack of coherence are only important when the inverse size of the target is smaller than the DM escape energy in the galaxy and for heavy DM particles.\footnote{For instance, for the Earth they are negligible for oxygen and only modest for iron, at mass resonance, whereas for the Sun and $\mx \sim 15$~GeV, they suppress capture by iron, but not significantly by oxygen or helium~\cite{Gould:1987ir}. Larger celestial bodies have higher escape velocities, so the suppression would affect more to lighter elements. Nevertheless, in this work we are interested in the minimum mass of DM particles that can be captured, for which these effects are less important than for higher masses. In any case, they are included below in our numerical computations and just neglected in this section for illustrative purposes.} Therefore, for the illustrative purposes of this section, we consider the zero-temperature limit and do not include the nuclear form factor. In this limit, the capture rate in the optically thin regime ($s_{\rm cap}(r) = 1$) reads~\cite{Gould:1987ir}
\begin{equation}
\mathcal{C}_{\rm weak} = \left(\frac{\rho_\chi}{\mx}\right) \langle v \rangle_0 \sum_i N_i \, \sigma_i \, 
\bigg\langle \frac{\hat{\phi}}{\langle \hat{\phi}\rangle_i} \, \left(1 - \frac{1 - e^{-\mathcal{B}_i^2}}{\mathcal{B}_i^2}\right) \, \xi_{\eta}(\mathcal{B}_i)\bigg\rangle_i \, \left(\frac{3}{2} \frac{v_e^2(R)}{v_d^2} \langle \hat{\phi}\rangle_i \right)  ~,
\end{equation}
where $\langle v \rangle_0 = \sqrt{8/(3\pi)} \, v_d$ is the average speed in the galactic rest frame, which depends on the position in the halo, and
\begin{equation}
\mathcal{B}_i^2(r) = \frac{3}{2} \frac{v_e^2(r)}{v_d^2} \frac{\mu_i}{\mu_{-,i}^2} \hspace{1cm} ; \hspace{3mm} 
\hat{\phi}(r) = \frac{v_e^2(r)}{v_e^2(R)} \hspace{1cm} ; \hspace{3mm} 
\langle \hat{\phi} \rangle_i = \frac{\int_0^R \hat{\phi}(r) \, n_i(r) \, 4 \pi \, r^2 \, \dd r}{N_i}  ~, 
\end{equation}
with $\mu_i = \mx/m_i$ and $\mu_{-,i} = (\mu_i -1)/2$. The term $\xi_{\eta}(\mathcal{B}_i)$ represents a suppression factor that accounts for the motion of the celestial body with respect to the halo frame ($\xi_0(\mathcal{B}_i) = 1$), with $\eta^2 = 3 \, v_{\rm cb}^2/(2 \, v_d^2)$~\cite{Gould:1987ir}. This is a consequence of a higher DM kinetic energy in the object's frame. For $\eta = 1$, $\xi_1(\mathcal{B}_i)$ monotonically takes values in the interval $(0.37-0.75)$, with the extremes reached for $\mathcal{B}_i^2(r) \ll 1$ and $\mathcal{B}_i^2(r) \gg 1$, respectively. Note also that $\xi_{\eta}(\mathcal{B}_i)$ includes a dependence on $\mathcal{B}_i$, but it is a ratio of two sums over all elements, so $\xi_{\eta}$ does not itself run over $i$, although it does depend on the position within the celestial body.

In the thick regime (large cross section), the geometric limit for the capture rate is usually considered~\cite{Bottino:2002pd, Bernal:2012qh, Garani:2017jcj}. This is an upper limit obtained from purely geometrical arguments and assumes that both, the probability of interaction and the probability of DM capture are one. This is, however, a non-physical limit, as even if the interaction probability is one, the probability of DM particles being captured (ending up with speeds below the escape velocity) is dictated by kinematics and is always smaller than one. This is corrected by the so-called saturation limit~\cite{Busoni:2017mhe}. In the limit of high escape velocities, $v_e > v_d$, the geometric limit only overestimates the saturation value by ${\cal O}(10\%)$ (for masses around the DM evaporation mass), in the case of the Sun, so it represents a reasonable approximation.\footnote{In the limit $v_e \to \infty$, the geometric and saturation limits coincide, as could be expected.} Nevertheless, when $v_e < v_d$, the geometric limit can grossly overestimate the maximum value for the capture rate (except for DM masses closely matching the targets mass). The overestimation scales as $v_d^4/v_e^4$ and therefore can be of several orders of magnitude, being more important for smaller objects.\footnote{This factor has been incorrectly neglected in some recent works that considered the maximum value of the capture rate~\cite{Neufeld:2018slx, Leane:2020wob, Pospelov:2020ktu, Leane:2021tjj}. Therefore, some of the presented results must be re-scaled, changing significantly the conclusions in some cases.} Although, the impact on the DM evaporation mass is generically mild, we use the more accurate saturation limit in what follows. In analogy to the capture rate for weak cross sections, the saturation limit can be written as 
\begin{equation}
\label{eq:satcapture}	
\mathcal{C}_{\rm sat} = \frac{\pi \, R^2}{\sum_i N_i \, \sigma_i} \left(\frac{\rho_\chi}{\mx}\right) \langle v \rangle_0 
\left[ \sum_i N_i \, \sigma_i \,  
\left(1 - \frac{1 - e^{-\mathcal{B}_i^2(R)}}{\mathcal{B}_i^2(R)}\right) \right] \, \xi_{1}(\mathcal{B}_i(R)) \, \left(\frac{3}{2} \frac{v_e^2(R)}{v_d^2} \right)  ~.
\end{equation}
In this section, for the sake of the discussion we consider this limit with a single element.

\subsection{DM annihilations in celestial bodies}

After DM particles get captured, further scatterings with the target elements, which are assumed to be in local thermodynamic equilibrium, would approximately thermalize them at a temperature $\tx(r)$ and attain a velocity distribution that can be approximated as Maxwell-Boltzmann.\footnote{For not too small cross sections, the thermalization time is typically much shorter than any relevant scale in this problem~\cite{Bertoni:2013bsa, Widmark:2017yvd, Liang:2018cjn, Blennow:2018xwu, Gaidau:2018yws, Garani:2020wge} (otherwise, the entire calculation would not be valid), such that the instantaneous approximation is adequate. Note also that the assumptions of a uniform and locally isotropic Maxwell-Boltzmann distribution for DM particles do not exactly hold in a realistic situation~\cite{Gould:1987ju, Gould:1989ez, Liang:2016yjf, Widmark:2017yvd}, although it is a reasonable approximation. Moreover, we assume the distribution to be locally truncated at the escape velocity, $v_e(r)$~\cite{Gould:1989hm}.} In the case of weak cross sections (optically thin regime), the DM radial distribution is approximately isothermal~\cite{Spergel:1984re, Faulkner:1985rm, Griest:1986yu},
\begin{equation}
\label{eq:DMdistiso}	
n_{\chi, {\rm iso}} (r, t) = N_\chi(t) \, \frac{e^{-\mx \phi(r)/\tx}}{\int_{0}^{R}  e^{-\mx \phi(r)/\tx} \, 4\pi r^2 \, \dd r} ~,
\end{equation}
where $\phi(r) = \int_0^r G M(r')/{r'}^2 \, \dd r'$, with $G$ the gravitational constant. For all DM masses, the DM temperature is a fraction of the central temperature of the object, being higher for heavier DM particles, which are more centrally concentrated. We compute the DM temperature, $\tx$, following Ref.~\cite{Garani:2017jcj}. In the case of large cross sections (optically thick regime), DM particles would thermalize locally with the medium and thus, $\tx(r) = T(r)$, with a radial distribution that can be approximated as~\cite{Nauenberg:1986em, Gould:1989hm}
\begin{equation}
\label{wq:DMdistLTE}
n_{\chi, {\rm LTE}} (r,t) = n_{\chi, {\rm LTE,0}}(t) \, \left(\frac{T(r)}{T(0)}\right)^{3/2} \, {\rm exp}\left(-\int_{0}^{r} \frac{\alpha(r') \frac{\dd T(r',t)}{\dd r'} + \mx \frac{\dd \phi(r')}{\dd r'}}{T(r')} \, \dd r'\right) ~.
\end{equation}
Here, $\alpha$ is the thermal diffusivity~\cite{Gould:1989hm, Gould:1990} and $n_{\chi, {\rm LTE},0}(t)$ is set by the normalization $\int_{0}^{R} n_{\chi, {\rm LTE}} (r) \, 4\pi r^2 \, \dd r = N_\chi(t)$. The transition between the thin and thick regimes can be described in terms of the Knudsen number, $\mathcal{K}$, which is defined as the ratio of the mean free path to the scale radius. Here, we follow Refs.~\cite{Bottino:2002pd, Scott:2008ns, Garani:2017jcj} to interpolate between the two regimes, using $\mathcal{K}_0 = 0.4$ as the pivot point~\cite{Gould:1989hm} for all cases. Note, however, that $\mathcal{K}$ depends on the properties of the capturing body. 

The radial distribution of thermalized DM particles clusters around the center of the capturing body, being more centrally concentrated in the case of heavier DM particles. Indeed, for all celestial bodies considered here, for the DM evaporation mass (i.e., for the lightest DM particles that can get efficiently captured), the scale radius of the distribution is $r_s \lesssim 0.2 \, R$. For larger objects, like stars, $r_s \lesssim 0.1 \, R$. Therefore, for the case of $s$--wave DM annihilations, which is considered throughout this work, the annihilation rate is given by
\begin{equation}
\label{eq:annihilationrate}
\mathcal{A} =  \langle \sigma_A v_{\chi \chi}\rangle \, \frac{\int_0^{R} n_\chi^2(r,t) \, 4 \pi \, r^2
		\, \dd r }{\left(\int_0^{R} \, n_\chi(r,t) \, 4\pi \, r^2 \, \dd r \right)^2}  \simeq \frac{\langle \sigma_A v_{\chi \chi}\rangle}{V_s} ~,
\end{equation}
where $\langle \sigma_A v_{\chi \chi} \rangle$ is the velocity-averaged DM annihilation cross section times the relative velocity of two DM particles, and throughout this paper we use the canonical value $\langle \sigma_A v_{\chi \chi} \rangle = 3 \times 10^{-26}~\textrm{cm}^3/\textrm{s}$ as our default value. In the second equality we have simply substituted the effective volume of integration by the volume at the scale radius, $V_s = 4/3 \, \pi \, r_s^3$, where most of the DM particles are concentrated. One can think of a region of approximately constant density and temperature (those at the core), which is a reasonable approximation within this small volume. In this section, we take $r_s = 0.1 \, R$ for illustrative purposes, although a more precise value affects our estimates in a negligible way.

\subsection{DM evaporation off celestial bodies}

In addition to the disappearance of captured DM particles due to annihilations, scatterings off targets of the medium could boost these particles to speeds above the local escape velocity, $v_e(r)$, such that they become gravitationally unbound. This process is referred to as DM evaporation, which is a finite-temperature process that depends on the temperature of the thermal bath in the celestial body that can transfer kinetic energy to DM particles. In the limit of zero temperature, there is no DM evaporation, as target particles cannot impinge any extra energy to thermalized DM particles. At finite temperature, energy conservation results in a higher probability for light DM particles to end up with higher final speeds, after scattering off thermal targets.  Thus, for small DM masses, this process is very efficient and sets a minimum mass of DM particles that can remain trapped in a celestial body. Lighter DM particles would get kicked out as they get captured. The evaporation rate is given by~\cite{Gould:1987ju}
\begin{equation}
\label{eq:ev+}
\mathcal{E} = \sum_i \int_0^{R} s_{\rm evap}(r) \, n_\chi(r,t) \, 4 \pi \, r^2 \, \dd r \, \int_0^{v_e(r)} f_\chi(\boldsymbol{w}, r) \, 4 \pi \, w^2 \, \dd w \, \int_{v_e(r)}^{\infty} \mathcal{R}_i^+ (w \rightarrow v) \, \dd v ~.
\end{equation}
where $f_\chi(\boldsymbol{w}, r)$ is the thermal velocity distribution of DM particles and $s_{\rm evap}(r)$ is a suppression factor that accounts for the fraction of DM particles that, even with a speed higher than the escape velocity, would actually escape due to further scatterings on their way out of the celestial body~\cite{Gould:1990}. We follow Refs.~\cite{Bernal:2012qh, Garani:2017jcj} to implement this suppression factor. As mentioned above, in the thin regime (long mean free path, $s_{\rm evap} \sim 1$), the DM distribution can be approximated as isothermal, with a temperature close to the central temperature of the celestial body. Thus, evaporation takes places mainly in a small region around the core. For large cross sections (thick regime, $s_{\rm evap} \ll 1$) the shell that contributes most to the evaporation rate moves towards the surface of the object, although it never reaches the last scattering surface (i.e., that for which the optical depth is equal to one)~\cite{Gould:1990}.

In the limit of the thin regime, $s_{\rm evap}(r) = 1$, an analytical solution for the evaporation rate per unit volume exists~\cite{Gould:1987ju}. For $\mx = m_i$, in the limit $E_e = \mx \, v_e^2/2 \gg T$, the solution is rather simple. Given that the DM evaporation mass is typically of the same order of the targets mass and the DM escape energy is larger than the thermal energy, we can consider this solution to illustrate the main features of the evaporation rate, which can be approximated as~\cite{Griest:1986yu, Gould:1987ju}
\begin{equation}
\mathcal{E} \simeq \sum_i \left[ \frac{1}{V_s} \frac{2}{\sqrt{\pi}} \left(\frac{2 \, \tx}{\mx}\right)^{1/2} \, \left(\frac{E_c}{\tx}\right) \, e^{-E_c/\tx} \, \right] \, N_i(r_{0.95}) \, \sigma_i ~,
\label{eq:evapapprox}
\end{equation}
where $E_c = \mx v_e^2(r=0)/2$ is the escape energy of DM particles at the core of the celestial body and $N_i(r_{0.95})$ is the number of targets $i$ within a radius $r_{0.95}$ such that $T(r_{0.95}) = 0.95 \, \tx$, which typically represents a small fraction of the total mass of the object. For the sake of illustration, in the next subsection we simply consider $N_i(r_{0.95}) = 0.1 \, M/m_i$, which is approximately correct for the Sun~\cite{Gould:1987ju}. The chosen value affects very little the calculation of the DM evaporation mass, because the evaporation rate depends exponentially on the DM mass. Moreover, note that Eq.~(\ref{eq:evapapprox}) is obtained for $\mu_i =1$ ($\mx = m_i$), but the evaporation rate depends very weakly on this ratio, except if $m_i \gg \mx$~\cite{Gould:1987ju}. Yet, in general, evaporation becomes efficient for $\mx$ not very different from $m_i$.

For large cross sections (thick regime), an analogous approximate expression for the evaporation rate can be written~\cite{Gould:1990}, including a suppression factor which accounts for the short mean free path of kicked DM particles. The expression, though, depends more critically on the temperature and density profiles of the celestial body than in the thin regime, and we do not explicitly consider it in this section, although we do compute it in detail for our numerical results. In any case, the discussion is qualitatively analogous.

\subsection{DM evaporation mass: a tale of two tails}

With all the relevant quantities briefly discussed and the simple approximations established, we turn to the estimation of the DM evaporation mass for a generic object. This is well known for the Sun and we simply restate and explain the results in the context of a generic celestial body with mass $M$, radius $R$ and core temperature $T_c$, whose properties remain approximately constant in time. We stress that the inputs used in this section are just presented for illustrative purposes, but we use the complete expressions to obtain our numerical results in Section~\ref{sec:results}.

We consider the saturation value for the capture rate, Eq.~(\ref{eq:satcapture}), and the annihilation and evaporation rates given in Eqs.~(\ref{eq:annihilationrate}) and~(\ref{eq:evapapprox}), respectively. Before equilibration ($t \ll \tau_{\rm eq}$), the DM evaporation mass grows with time ($\mathcal{E} \, t \simeq \ln(11)$) up to its maximum value at equilibrium. At this point, the DM evaporation mass is independent of time (assuming all the properties of the capturing object remain the same) and, from Eq.~(\ref{eq:evapmass}), can be defined as~\cite{Garani:2017jcj}
\begin{equation}
\mathcal{E} (m_{\rm evap}) \, \tau_{\rm eq}(m_{\rm evap}) \simeq \frac{1}{\sqrt{0.11}} ~.
\end{equation}
For $3 \, v_e^2 \, \mu_i \gg 2 \, v_d^2 \, \mu_{-,i}^2$ ($\mathcal{B}_i^2 \gg 1$), representative of stars like the Sun (except for heavy DM, which is not the focus of this paper), and for the geometric cross section, $\sum N_i \, \sigma_i^{\rm geom} = \pi \, R^2$, which approximately sets the maximum capture rate (the saturation value), the equilibration time approximately scales as (using $r_s = 0.1 \, R$)
\begin{equation}
\label{eq:taustars}	
\tau_{\rm eq} \simeq  7 \times 10^{12}~{\rm s} \, \left(\frac{M_\odot}{M}\right)^{1/2} \, \left(\frac{R}{R_\odot}\right) \, \left(\frac{0.4~\textrm{GeV/cm}^3}{\rho_\chi}\right)^{1/2} \, \left(\frac{\mx}{\textrm{GeV}}\right)^{1/2} \, \left(\frac{v_d}{270~\textrm{km/s}}\right)^{1/2} \, \left(\frac{3 \times 10^{-26}~\textrm{cm}^3/\textrm{s}}{\langle \sigma_A v_{\chi\chi} \rangle}\right)^{1/2} ~.
\end{equation}
The evaporation rate approximately scales as (using $\sigma_{\rm geom} \, N(r_{0.95}) = 0.1 \, \pi \, R^2$)
\begin{equation} 
\mathcal{E} \simeq 0.06~\textrm{s}^{-1} \, \left(\frac{R_\odot}{R}\right) \, \left(\frac{\tx}{1.5 \times 10^7~\textrm{K}}\right)^{1/2}\, \left(\frac{\textrm{GeV}}{\mx}\right)^{1/2} \, \left(\frac{E_c}{\tx}\right) \, e^{-E_c/\tx} ~,	
\end{equation}	
where we have kept the dependence on $E_c/\tx$ explicit. For celestial bodies like the Sun (and in general for other objects), the prefactor is very large, so it is obvious that $E_c/\tx \gg 1$ in order to suppress the evaporation rate at the level of $\tau_{\rm eq}^{-1}$. Putting these two quantities together, the equation for the DM evaporation mass reads
\begin{equation}
\label{eq:eqmevapstars}
\left(\frac{E_c}{\tx}\right) e^{-E_c/\tx} \simeq 7 \times 10^{-12} \, \left( \frac{M}{M_\odot}\right)^{1/2} \left(\frac{1.5 \times 10^7~\textrm{K}}{\tx}\right) ^{1/2}  \left( \frac{\rho_\chi}{0.4~\textrm{GeV/cm}^3}\right)^{1/2} \left(\frac{270~\textrm{km/s}}{v_d} \right)^{1/2} \left(\frac{\langle \sigma_A v_{\chi \chi}\rangle}{3 \times 10^{-26}~\textrm{cm}^3/\textrm{s}}\right)^{1/2} ~.
\end{equation}
For the representative (solar) values used in the above equation, the solution is $E_c/\tx \simeq 29$, which results in a DM evaporation mass for the Sun, $m_{\rm evap} \simeq 3.2$~GeV, using $v_e^2(r=0) = 5 \, v_e^2(r=R)$ and $\tx = 0.9 \, T_c$. This is known in the literature for over three decades~\cite{Spergel:1984re, Gaisser:1986ha, Griest:1986yu, Gould:1987ju}. Already the authors of Ref.~\cite{Griest:1986yu} explicitly wrote that the relevant value to compute the DM evaporation mass for the Sun is $E_c/\tx \simeq 30$. Furthermore, note that for smaller scattering cross sections, the right-hand side scales as $\sigma^{-1/2}$, so the DM evaporation mass is slightly smaller. For larger cross sections, there is an extra (exponential) suppression term in the evaporation rate (left-hand side in the above equation) and the DM evaporation mass is also smaller, and more pronouncedly than for smaller cross sections. Notice also that the details of the density and temperature profiles are embedded in $v_e^2(0)$ and $T_\chi$, which cannot vary much given the mass, radius and core temperature of the celestial body, and thus, affect little the value of the DM evaporation mass. In particular, variations on $v_e^2(0)$ are expected to be $\lesssim 10\%$, which also applies to the smaller objects we discuss next.

In the opposite limit, representative of the Earth (and other solar system planets), $\mathcal{B}_i^2 \ll 1$. Therefore, in addition to the $v_e^2/v_d^2$ suppression, there is an extra suppression factor in the capture rate, $\mathcal{B}_i^2/2 = (3 \, v_e^2 \, \mu_i) / (4 \, v_d^2 \, \mu_{-,i}^2)$, except in a very narrow mass range, where the mass of the targets and of the DM particles closely match (in that case, the results for $\mathcal{B}_i \gg 1$ are reproduced). This implies a longer equilibration time and therefore, a higher DM evaporation mass. In this limit, the equations for the equilibration time, the evaporation rate and the DM evaporation mass read
\begin{eqnarray}
\label{eq:tauplanets}		
\tau_{\rm eq} & \simeq & 1.5 \times 10^{15}~{\rm s} \, \left(\frac{\mu_{-}^2}{3 \, \mu}\right)^{1/2} \left(\frac{M_\oplus}{M}\right) \left(\frac{R}{R_\oplus}\right)^{3/2} \\
& & \hspace{1.5cm} \times \left(\frac{0.4~\textrm{GeV/cm}^3}{\rho_\chi}\right)^{1/2} \left(\frac{\mx}{\textrm{GeV}}\right)^{1/2} \left(\frac{v_d}{270~\textrm{km/s}}\right)^{3/2} \left(\frac{3 \times 10^{-26}~\textrm{cm}^3/\textrm{s}}{\langle \sigma_A v_{\chi\chi} \rangle}\right)^{1/2} ~, \nonumber \\[2ex]
\mathcal{E} & \simeq & 0.1~\textrm{s}^{-1} \, \left(\frac{R_\oplus}{R}\right)  \left(\frac{\tx}{6000~\textrm{K}}\right)^{1/2} \left(\frac{\mx}{\textrm{GeV}}\right)^{-1/2}  \left(\frac{E_c}{\tx}\right) \, e^{-E_c/\tx} ~, \\[2ex]
\left(\frac{E_c}{\tx}\right) \, e^{-E_c/\tx} 
& \simeq & 2 \times 10^{-14} \, \left(\frac{3 \, \mu}{\mu_{-}^2}\right)^{1/2} \left( \frac{M}{M_\oplus}\right)  \left(\frac{R_\oplus}{R}\right)^{1/2} \left(\frac{6000~\textrm{K}}{\tx}\right)^{1/2} \label{eq:eqmevapplanets} \\
& & \hspace{1.2cm} \times \left(\frac{\rho_\chi}{0.4~\textrm{GeV/cm}^3}\right)^{1/2} \left(\frac{270~\textrm{km/s}}{v_d}\right)^{3/2} \left(\frac{\langle \sigma_A v_{\chi \chi}\rangle}{3 \times 10^{-26}~\textrm{cm}^3/\textrm{s}}\right)^{1/2} ~, \nonumber
\end{eqnarray}
which results in $E_c/\tx \simeq 34$ for the Earth, and in a DM evaporation mass $m_{\rm evap} \simeq 13$~GeV, where we have used $v_e^2(r=0) = 1.9 \, v_e^2(r=R)$, $\tx = T_c$ and $\mu =1/3$. Note that $E_c/\tx$ is similar to the value obtained in the limit $\mathcal{B}_i^2 \gg 1$ and $m_{\rm evap}$ is very close to the value quoted in Ref.~\cite{Gould:1988eq}. Indeed, the fact that the DM evaporation mass is approximately given by $E_c/\tx \sim 30$ is a robust result that applies to all round objects in hydrostatic equilibrium. The same considerations regarding the dependence on the scattering cross section can be drawn for $\mathcal{B}_i^2 \ll 1$. In this limit, however, the equilibration time is longer. Yet, this is approximately the shortest possible equilibration time (we are using a capture rate close to maximum), so cross sections smaller than the geometric value could result in equilibration times longer than the age of the object. In those cases, the approximate equation to be used is $\mathcal{E} \, t \simeq \ln(11)$ and the DM evaporation mass grows with time. Indeed, it is unlikely the smallest objects we consider reach equilibrium, unless the DM mass closely matches the mass of some of the targets or they are close to the center of their host halo (high $\rho_\chi$ and low $v_d$).

All in all, the exponential dependence of the evaporation rate, which sets the DM evaporation mass along the exponential tail, is really a tale of two tails~\cite{Gaisser:1986ha}. On one hand, a DM particle in the high-velocity tail of its distribution could scatter off a target with typical thermal speed, and be promoted to a speed higher than the escape velocity. On the other hand, a DM particle with typical thermal speed may be kicked off the celestial object due to the scattering with a target in the high-velocity tail of its distribution. The first process is the most important one~\cite{Gaisser:1986ha}.

\begin{figure}[t]
	\centering
	\includegraphics[width=0.9\linewidth]{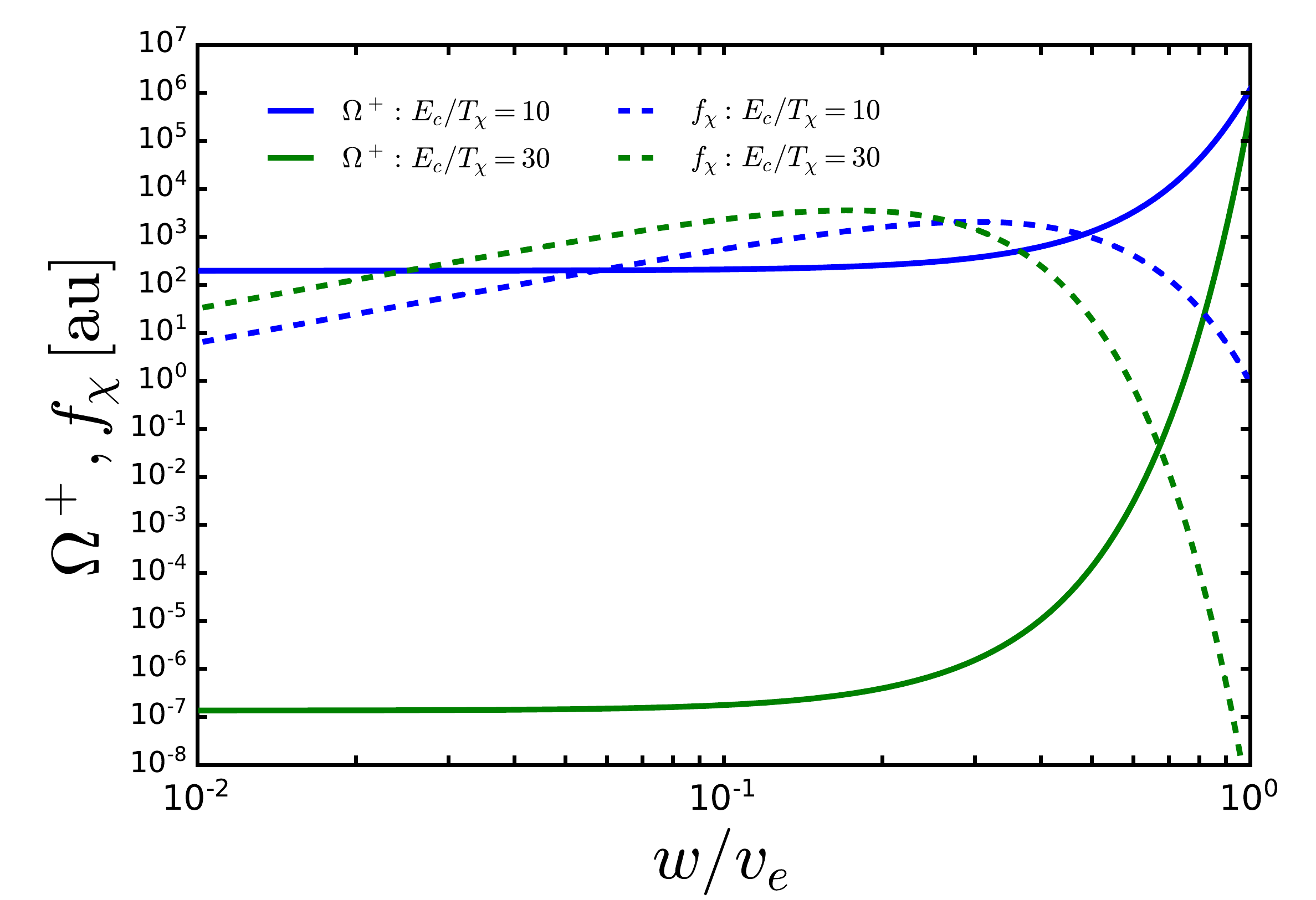}
	\caption{\textbf{\textit{The two tails of the evaporation rate}}. The probability for a DM particle with speed $w$ to scatter off a target in the medium, with temperature $T$, and gain enough energy to escape from the gravitational potential of the capturing object, $\Omega_{v_e}^+(w)$ (solid lines), and the velocity distribution of DM particles, with temperature $\tx = 0.9 \, T$, $f_\chi(w)$ (dashed lines), both in arbitrary units. We show these distributions for $E_c/\tx = 10$ (blue lines) and $E_c/\tx = 30$ (green lines), using $\mu = 1$.}  	
	\label{fig:tails}
\end{figure}

One way to visualize the exponential suppression of the evaporation rate is by plotting the thermal velocity distribution of DM particles (with temperature $\tx$), which scales as $\sim \left(E_e/\tx\right)^{3/2} (w/v_e)^2 \, \exp[- (E_e/\tx) \, (w/v_e)^2]$, and the probability for a DM particle with speed $w$ to interact with a target particle (with a thermal distribution of temperature $T(r)$) and gain enough energy to escape. The latter probability is proportional to $\Omega_{v_e}^+(w) = \int_{v_e}^\infty \mathcal{R}^+(w \to v) \, \dd v$~\cite{Gould:1987ju}, and approximately scales as $\sim \exp[- (E_e/T) (1 - (w/v_e)^2)]$. We evaluate $E_e$ at the core, $E_c$, because in the thin regime DM evaporation mostly occurs close to the center of the celestial body. In Fig.~\ref{fig:tails} we show both distributions as a function of $w/v_e$, for two values of $E_c/\tx$. As can be clearly seen, the smaller $E_c/\tx$, the larger the overlap of the two distributions, or in other words, the higher the evaporation rate. For small $E_c/\tx$, more DM particles have speeds close to the escape velocity (which after evaporation get repopulated by the thermalization process), so they need less energy to evaporate. Moreover, the smaller $E_c/\tx$, the higher the probability for DM particles to end up with speeds higher than the escape velocity. Therefore, the overall probability for this to happen is higher, and in relative terms, the ratio of probabilities for two values of $E_c/\tx$ approximately scales as the exponential of the difference between the two values of $E_c/\tx$. These two effects go in the same direction and result in a huge evaporation rate when a significant overlap occurs. As a consequence, in order to suppress sufficiently the evaporation rate to maintain an equilibrium population of DM particles, the two distributions should only overlap far in their exponential tails and $E_c/\tx \sim 30$ is generically required. We stress that this is a robust result, which not only applies to the Sun, but to all round celestial bodies in hydrostatic equilibrium, and provides the correct result, for constant scattering cross sections at the geometric value, within $\lesssim 30\%$ accuracy.

Finally, note that some recent works, which also assumed constant scattering cross sections, did not follow these arguments and incorrectly estimated the DM evaporation mass~\cite{Bramante:2019fhi, Ilie:2020nzp, Leane:2020wob, Leane:2021ihh, Leane:2021tjj}. In Ref.~\cite{Bramante:2019fhi}, the DM evaporation mass was defined as the DM mass for which the thermal radius (obtained from the virial theorem) is equal to the radius of the capturing body, which is similar to setting $E_e/\tx \sim 1$. This implies a DM evaporation mass for the Earth of $\sim 100$~MeV, which is not correct. For larger masses with smaller thermal radius than the radius of the object, the evaporation rate is already very high and, as explained above, those DM particles would also evaporate. Under the very same assumptions, the correct value of the DM evaporation mass for the Earth is $m_{\rm evap} \sim 12$~GeV~\cite{Freese:1985qw, Krauss:1985aaa, Gould:1988eq, Garani:2019rcb}. On the other hand, Ref.~\cite{Leane:2020wob} proposes the observation of exoplanets and brown dwarfs to search for effects of captured DM particles with sub-GeV mass. Nevertheless, these authors set the condition for DM particles to remain trapped to $E_e/\tx > 1$,\footnote{Actually, the authors of Ref.~\cite{Leane:2020wob} use the condition $3 \, T(r)/2 < G M(r) \, \mx/(2 \, r)$, which, as argued above, is not correct. Additionally, the right-hand side is claimed to be the gravitational potential at position $r$, but notice that this not correct. Moreover, it is not clear how the minimum bounded DM masses which are quoted in that paper were obtained.} neglecting the crucial exponential tail of the evaporation rate. Again, this implies a gross underestimation of the DM evaporation mass and explains why Ref.~\cite{Leane:2020wob} (and also Refs.~\cite{Leane:2021ihh, Leane:2021tjj}) incorrectly claimed DM evaporation masses well in the sub-GeV regime. A DM evaporation mass for the Earth of $\sim 100$~MeV was also incorrectly estimated, as well as for other planets and brown dwarfs, whose DM evaporation mass was also greatly underestimated to be as low as $m_{\rm evap} \sim 4.5$~MeV.\footnote{In addition to neglecting the exponential tail of the evaporation rate, Refs.~\cite{Leane:2020wob, Leane:2021ihh} also consider core temperatures for brown dwarfs which are about a factor of six smaller than what we do (see below). Overall, this amounts to about a factor of $\sim 200$ underestimation of the DM evaporation mass with respect to our result.} A more accurate estimation for the most massive brown dwarfs was $m_{\rm evap} \sim {\cal O}(1)$~GeV~\cite{Zentner:2011wx}. Below, this is explicitly discussed in more detail and contrasted with our results.

\subsection{DM evaporation off compact bodies: white dwarfs and neutron stars}

As we illustrate below, for white dwarfs and neutron stars, the DM evaporation mass is much smaller than the targets masses, $\mu_i \ll 1$. Therefore, the approximation we considered above to estimate the evaporation rate (i.e., $E_e/T \gg 1$) is not appropriate in these cases, as for these objects, $\mu_i \, E_e/T \ll 1$. In addition, so far in this section, we have only considered the case of non-degenerate targets. This is valid for all cases discussed in this paper except for neutron stars (and for brown and white dwarfs if the targets are electrons, which we do not consider in this paper). In the case of neutron stars, the targets we study are degenerate neutrons and the calculation of the capture and evaporation rates of DM particles (and also of the DM thermalization time~\cite{Bertoni:2013bsa, Garani:2020wge}) needs to be modified to properly include the effect of Pauli blocking. This is approximately accounted for by adding a correction factor, which in the case of the evaporation rate is $\zeta_i \simeq {\rm min}\{2 \, T_c/p_{{\rm F}, i} \, ,1\}$~\cite{Garani:2018kkd}, where $p_{{\rm F}, i}$ is the Fermi momentum of target particles $i$. As will become obvious in the next paragraphs, for the case of interactions with nucleons, the correction factor in the capture rate is not relevant in the discussion of the DM evaporation mass.

Including the impact of Pauli blocking and taking the limit $\mu_i \, E_e/T \ll 1$ (with $E_e/T \gg 1$), the evaporation rate for equal temperatures ($T_c = \tx$), using Ref.~\cite{Gould:1987ju}, can be written as
\begin{equation}
\mathcal{E} \simeq \sum_i \, \frac{4}{3} \left(\frac{\mu_i}{\pi}\right)^{1/2} \, \left(\frac{E_c}{\tx}\right)^{1/2} \, \zeta_i \, \left[ \frac{1}{V_s} \frac{2}{\sqrt{\pi}} \left(\frac{2 \, \tx}{\mx}\right)^{1/2} \, \left(\frac{E_c}{\tx}\right) \, e^{-E_c/\tx} \, \right] \, N_i(r_{x}) \, \sigma_i ~.
\label{eq:evapWD}
\end{equation}
For non-relativistic and degenerate target particles, such as neutrons in neutron stars, the Fermi momentum is $p_{\rm F} \simeq \sqrt{2 \, m_i \, \mu_{\rm F}}$, with $\mu_{\rm F} \simeq 350$~MeV~\cite{Garani:2018kkd}. For relativistic and fully degenerate target particles, such as electrons in white dwarfs and in neutron stars, $p_{\rm F} \simeq \mu_{\rm F}$. As done above for other celestial bodies, for the following estimates we take $\sum_i N_i(r_x) \, \sigma_i = 0.1 \, \pi \, R^2$, although the DM evaporation mass is very little sensitive to its precise value. 

Now, we establish a distinction between white dwarfs and neutron stars. For the former, we just proceed as in the preceding subsection, but using Eq.~(\ref{eq:evapWD}) for the DM evaporation rate. For the latter, however, $m_{\rm evap} \lesssim 100$~keV, so the annihilation cross section cannot be taken as the canonical value (otherwise, the DM contribution would overclose the Universe!). Thus, for neutron stars, we assume a very small annihilation cross section, so that equilibration between capture and annihilation does not take place. Using Eq.~(\ref{eq:evapmass}) with $\mathcal{A} = 0$, the (time-dependent) DM evaporation mass is defined by $\mathcal{E}(m_{\rm evap}) \, t_{\rm NS} = \ln(11)$, where $t_{\rm NS}$ is the age of the neutron star. This is effectively equivalent to asymmetric DM scenarios.

Therefore, for white dwarfs, and considering only interactions with nuclei, as done throughout this work (i.e., $\zeta_i = 1$), the DM evaporation mass can be obtained by solving
\begin{equation}
\left(\frac{E_c}{\tx}\right)^{3/2} e^{-E_c/\tx} \simeq 6 \times 10^{-11} \, \left( \frac{M}{M_\odot}\right)^{1/2} \left(\frac{4 \times 10^5~\textrm{K}}{\tx}\right) ^{1/2}  \left( \frac{\rho_\chi}{0.4~\textrm{GeV/cm}^3}\right)^{1/2} \left(\frac{270~\textrm{km/s}}{v_d} \right)^{1/2} \left(\frac{\langle \sigma_A v_{\chi \chi}\rangle}{3 \times 10^{-26}~\textrm{cm}^3/\textrm{s}}\right)^{1/2} ~.
\end{equation}
For a white dwarf with $M = M_\odot$ ($R \simeq 0.9~R_\oplus$) and $T_c = \tx = 4 \times 10^5$~K, this results in $E_c/\tx \simeq 29$, which is in the same ballpark as the values for other objects. In this case, the DM evaporation mass is $m_{\rm evap} \simeq 1.0$~MeV, where we have used $v_e^2(r=0) = 3.9 \, v_e^2(r=R)$.

In the case of neutron stars, the equation for the DM evaporation mass, $\mathcal{E}(m_{\rm evap}) \, t_{\rm NS} = \ln(11)$, can be written as
\begin{equation}
	\left(\frac{E_c}{\tx}\right)^{3/2} e^{-E_c/\tx} \simeq 3 \times 10^{-12} \, \left(\frac{p_\textrm{F}}{0.8~\textrm{GeV}}\right) \, \left(\frac{10^5~\textrm{K}}{\tx}\right) ^{3/2}  \left(\frac{R}{11.5~\textrm{km}}\right) \left(\frac{4.5~\textrm{Gyr}}{t_{\rm NS}} \right)~.
\end{equation}
For a neutron star with $M = 1~M_\odot$, $R = 11.5$~km, $T_c = \tx = 10^5$~K and $t_{\rm NS} = 4.5$~Gyr, this results in $E_c/\tx \simeq 32$, again very similar to the values for other objects. Using $v_e^2(r=0) = 1.5 \, v_e^2(r=R)$, the DM evaporation mass is $m_{\rm evap} \simeq 1.4$~keV. 

Thus, the DM evaporation mass for white dwarfs and neutron stars is also approximately given by $E_c/\tx \sim 30$. We stress that this is a general and robust result which applies to all the objects we consider in this work, that is, to all spherical celestial bodies in hydrostatic equilibrium. In any case, these are approximate estimates, which allow us, nonetheless, to obtain the DM evaporation mass with a precision of a few tens of percent. We obtain our results in Section~\ref{sec:results} following Ref.~\cite{Garani:2018kkd} for the calculation of the evaporation rate in neutron stars for DM scattering off non-relativistic degenerate neutrons, and we follow the discussion above~\cite{Garani:2017jcj} for all other objects. Although for the calculation of DM evaporation mass in neutron stars a more accurate treatment must use relativistic kinematics~\cite{Bell:2020jou, Bell:2020lmm}, the required corrections do not significantly change the results obtained here.

\section{Main properties of celestial bodies}
\label{sec:celestialbodies}

\begin{figure}[t]
	\centering
\hspace*{-0.65cm}	\includegraphics[width=1.05\linewidth]{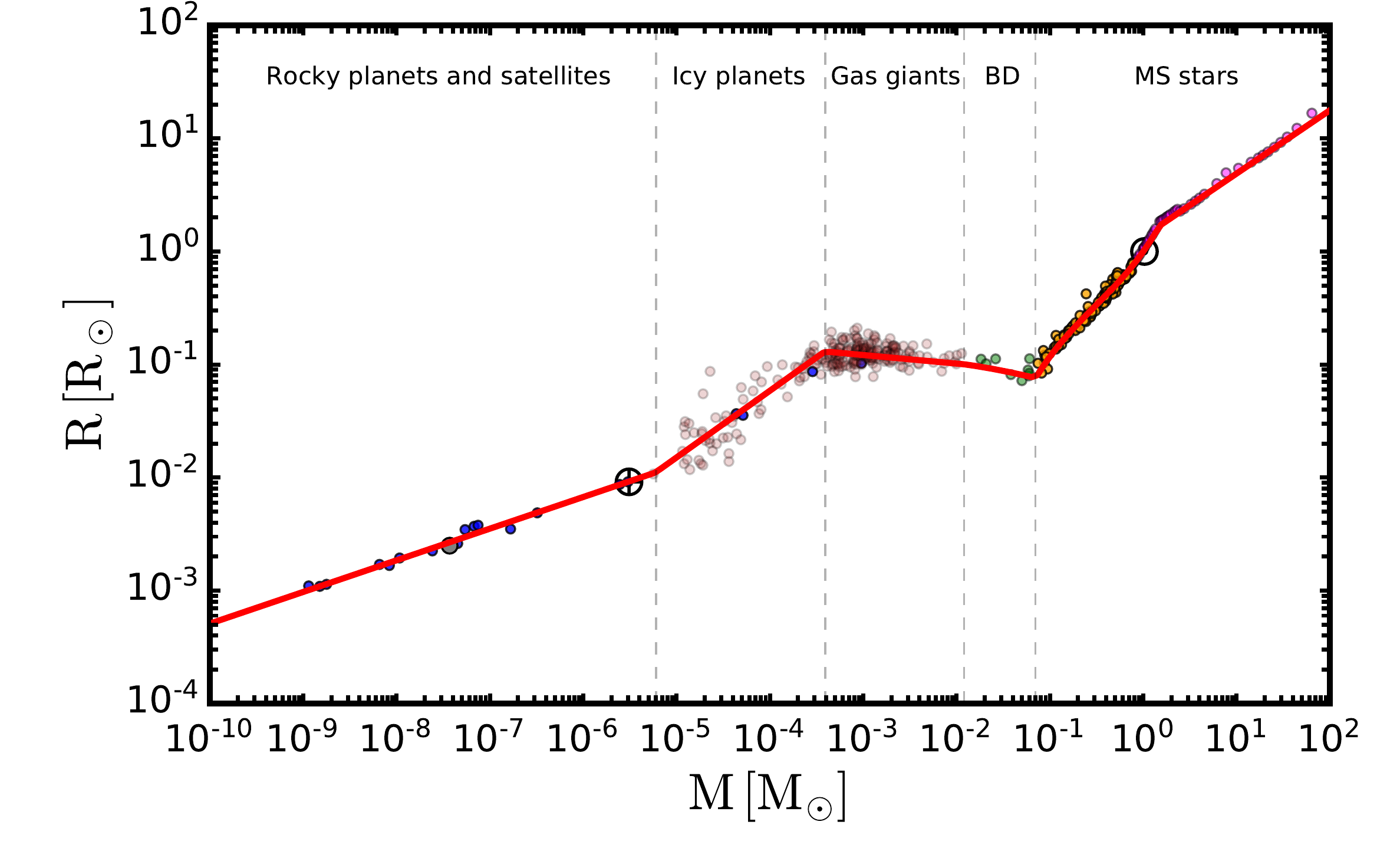}
	\caption{\textbf{\textit{Radius of planetary bodies, brown dwarfs and main-sequence stars}}, as a function of the mass of the object and using the mass--radius relation described in the text. We also show a compilation of data corresponding to planetary bodies~\cite{Chen:2017} (solar system planets and satellites in blue), brown dwarfs with measured radius~\cite{Deleuil:2008tg, Johnson:2010kx, Bouchy:2010iv, Anderson:2010qx, Siverd:2012fz, Diaz:2013jea, Moutou:2013nba, Littlefair:2014}, low-mass stars~\cite{Parsons:2018}, intermediate-mass and massive stars~\cite{Eker:2018}.}  	
	\label{fig:MRR}
\end{figure}

\begin{figure}[t]
	\centering
\hspace*{-0.65cm}	\includegraphics[width=1.05\linewidth]{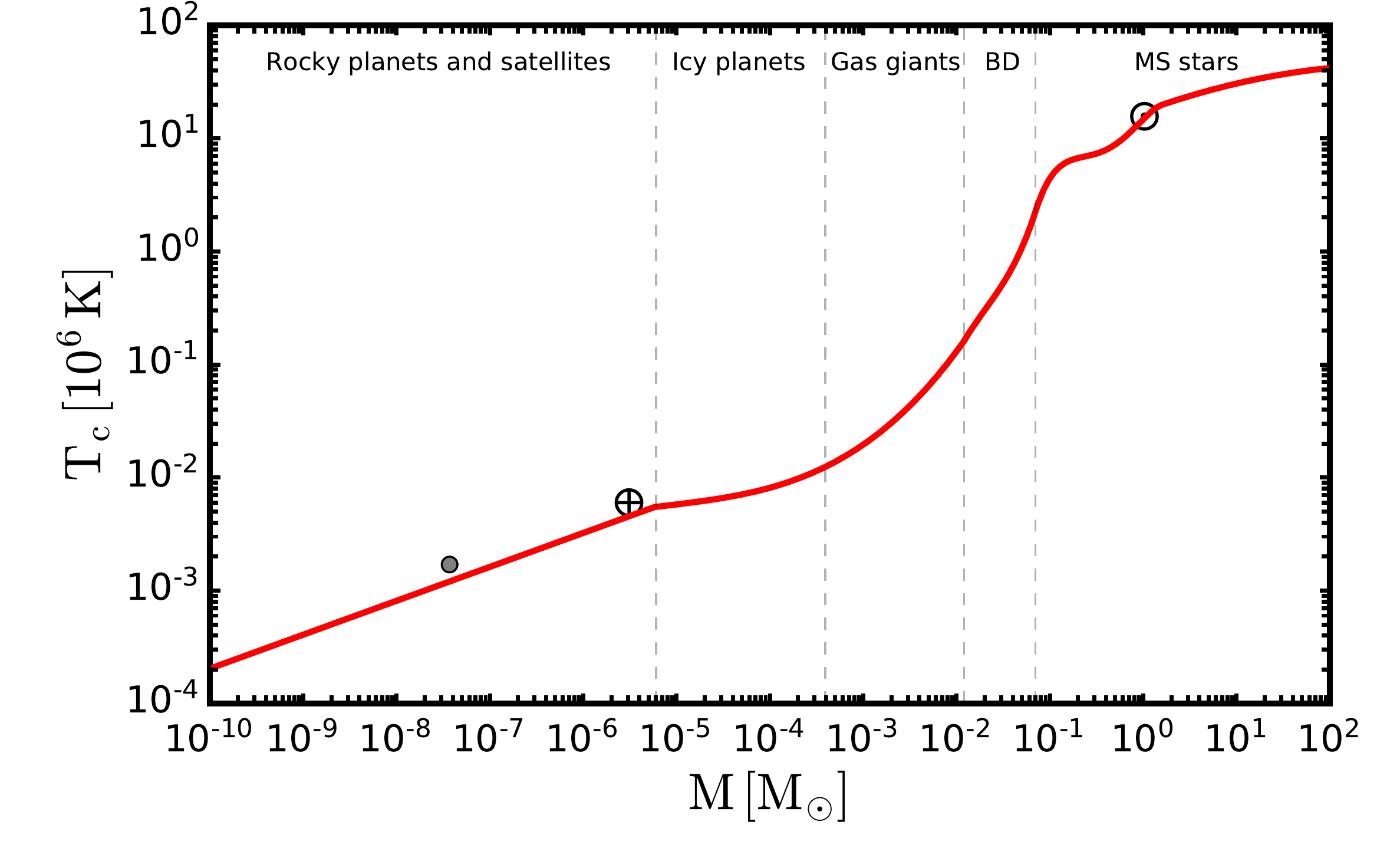}
	\caption{\textbf{\textit{Core temperature of planetary bodies, brown dwarfs and main-sequence stars}}, as a function of the mass of the object and using the mass--core temperature relation described in the text. From detailed models and data, the estimated core temperature for the Moon, Earth and Sun is also indicated.}
	\label{fig:MTR}
\end{figure}

\begin{figure}[t]
	\centering
\hspace*{-0.65cm}	\includegraphics[width=1.05\linewidth]{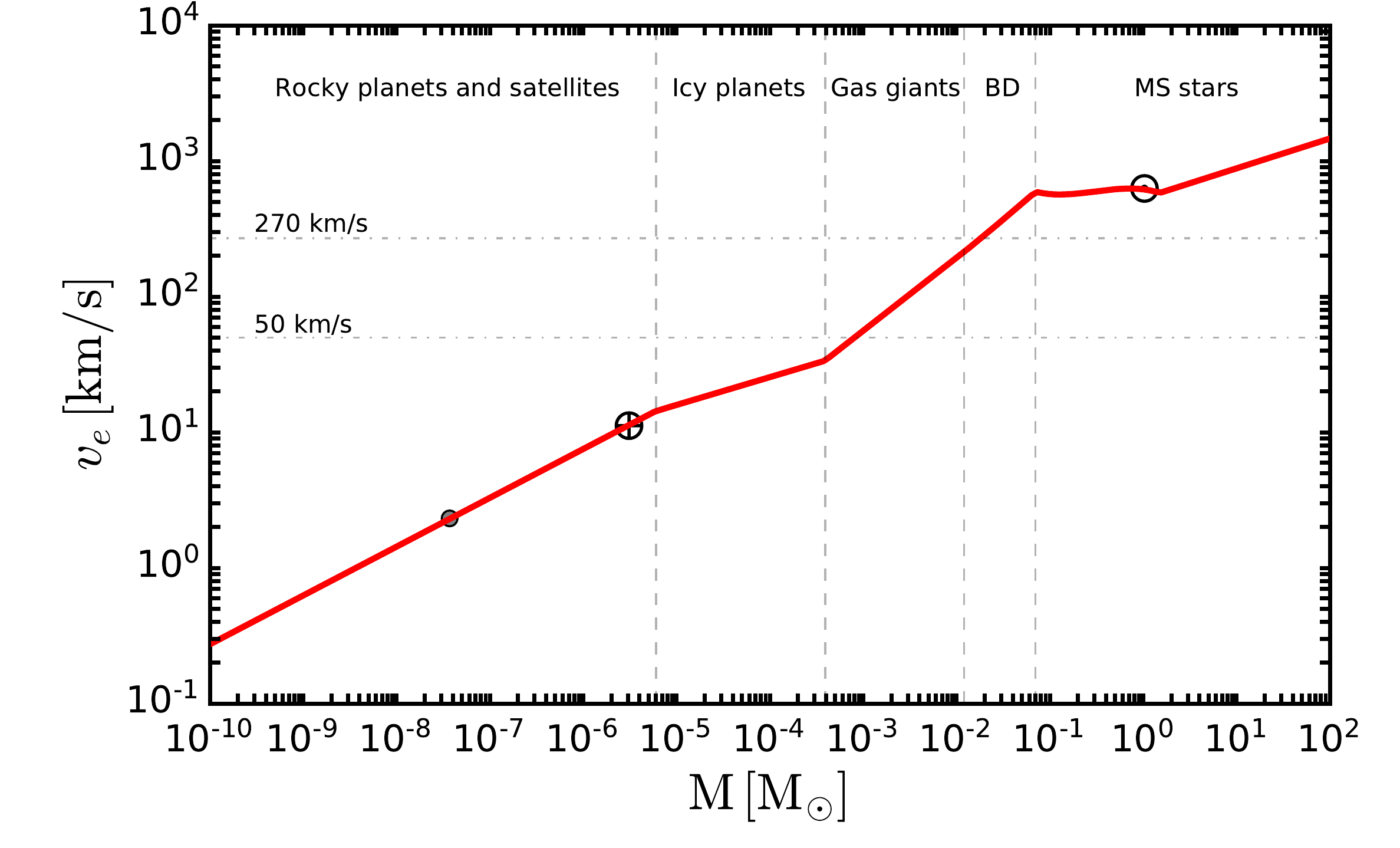}
	\caption{\textbf{\textit{Escape velocity at the surface of planetary bodies, brown dwarfs and main-sequence stars}}, as a function of the mass of the object and using the mass--radius relation reported in the text. The values for the Moon, the Earth and the Sun are shown. We also indicate two values of the galactic dispersion velocity $v_d$ (dashed lines), which are references for the local neighborhood and near the galactic center. The DM capture rate by objects with $v_e < v_d$ is suppressed by a factor proportional to $(v_e/v_d)^4$, but with resonance-like features for DM masses matching targets masses.}  	
	\label{fig:Mve}
\end{figure}

In this section, we describe the average properties of celestial round bodies in hydrostatic equilibrium, spanning a wide mass range, $[10^{-10} - 10^2]~M_\odot$. In order to determine the DM evaporation mass, the required main characteristics of the capturing objects are the mass $M$, radius $R$, and their density $\rho(r)$ and temperature $T(r)$ profiles. We stress that, for a given object's mass, radius and core temperature, the DM evaporation mass depends little on the shape of the density (mainly via the ratio of the gravitational potential at the center and at surface) or temperature profiles. The kinematics of elastic scattering depends on the mass of the DM particles as well as that of the targets, so another important factor is the composition of the material. For the sake of simplifying the discussion, we include hydrogen (in terms of the $X_{\rm H} \equiv X$ mass fraction), helium ($X_{\rm He} \equiv Y$ mass fraction) and as representative of heavier elements we consider carbon, oxygen, water, silicate perovskite (MgSiO$_3$) and iron (the mass fraction of these heavier elements is generically denoted by $Z$). 

Given that some generic features of celestial bodies can be approximately described in terms of polytropes, we first briefly introduce the properties of objects with this kind of equation of state. We shall later use them as ballpark models for some cases. Next, we provide an overview of the general properties of planetary objects, brown dwarfs, main-sequence stars, post-main-sequence evolutionary phases of stars, white dwarfs and neutron stars. The process of DM capture is assumed not to modify their properties in a significant way, so that we can still use results from standard modeling without including DM effects. 

Throughout this work, we consider the mass as the single variable that determines the rest of the properties of celestial bodies, in an average way. We provide parameterizations for the radius, core temperature, density and temperature profiles and composition, as a function of the mass of the object. All of them are based on actual data and modeling. We just impose continuity at the transitions from one mass range to another.

The mass--radius and mass--core temperature relations reported in this section are shown in Figs.~\ref{fig:MRR} and~\ref{fig:MTR}, respectively. The escape velocity at surface is shown as a function of the mass of celestial bodies in Fig.~\ref{fig:Mve}. For those objects located far enough from the center of the host halo such that their local dispersion velocity is higher than the escape velocity of the object, the capture rate is suppressed (except at mass-matching, where resonance-like features appear) and hence, the equilibration time is longer (see below) and the DM evaporation mass is larger (assuming equilibrium is reached).

\subsection{Polytropic models}

The interiors of some celestial bodies are reasonably well described by gases with polytropic equations of state, such that $P(r) = K \, \rho(r)^{1 + 1/n}$, where $P(r)$ is the pressure, $\rho(r)$ is the density, $K$ is a proportionality constant and $n$ is the polytropic index. This index evolves in mass from $n \gtrsim 0$, corresponding to rocky planets with Earth-like masses, to $n = 3$, corresponding to massive stars with a radiative core~\cite{Chabrier:2008bc}. For intermediate masses, from Jovian planets up to low-mass brown dwarfs, objects are well approximated by $n \simeq 1$, and brown dwarfs by $n \simeq 3/2$. This variation covers a range of about nine orders of magnitude in mass. The main advantage of this type of models is that pressure only depends on density, so only the hydrostatic and Poisson equations are needed, with no reference to heat transfer or thermal balance. Although this might seem an oversimplification, these models have proven to be remarkably useful in the interpretation of many features of the structure of celestial bodies and have already been used in the context of DM capture and evaporation in stars~\cite{Spergel:1984re, Spergel:1988, Dearborn:1990, Freese:2008wh, Frandsen:2010yj, Ilie:2020nzp}, so we consider them to obtain a generic description. Therefore, we first describe the distribution of density, pressure and temperature of polytropic models. 

We consider a celestial body with mass $M$ and radius $R$ constituted of a material with an equation of state of a polytrope of index $n$. The Poisson and hydrostatic equations (assuming spherical symmetry) can be written as
\begin{eqnarray}
\frac{1}{r^2}  \frac{\partial}{\partial r} \left(r^2 \, \frac{\partial \Phi}{\partial r}\right) & = & 4 \pi G \, \rho(r) ~, \label{eq:Poisson} \\[1ex]
\frac{\partial P}{\partial r} & = & - \frac{\partial \Phi}{\partial r} \, \rho(r) ~. 
\end{eqnarray}
By substituting the expression of the pressure in terms of the density for a polytrope, the hydrostatic equation can be integrated. Setting the gravitational potential $\Phi(R) = 0$ at the surface $\rho(R) = 0$, one gets
\begin{equation}
\rho(r) = \left( \frac{- \Phi(r)}{(n + 1) \, K}\right)^n ~.
\label{eq:hydrosol}
\end{equation}
Now, we introduce the dimensionless quantities $\theta_n$ and $\xi$, 
\begin{equation}
\theta_n = \frac{\Phi}{\Phi_c} = \left(\frac{\rho}{\rho_c}\right)^{1/n} \hspace{1cm} ; \hspace{1cm} 
\xi = \frac{r}{r_n} \hspace{5mm} , \hspace{5mm} r_n^2 = \frac{(n + 1)^n \, K^n}{4 \pi G \, (- \Phi_c)^{n - 1}} = \frac{(n + 1) \, K}{4 \pi G \, \rho_c^{1-1/n}} ~,
\label{eq:defpoly}
\end{equation}
where the subscript $_c$ indicates quantities evaluated at the center of the object. Note that $P(r)  = P_c \, \theta^{1+n}_n(r)$. Substituting Eq.~(\ref{eq:hydrosol}) into Eq.~(\ref{eq:Poisson}), in terms of the variables defined in Eq.~(\ref{eq:defpoly}), the Lane-Emden equation is obtained,
\begin{equation}
\frac{1}{\xi^2} \, \frac{d}{d \xi} \left(\xi^2 \, \frac{d \theta_n}{d \xi}\right) + \theta_n^n = 0 ~.
\end{equation}
Finite solutions require $d\theta_n(0)/d\xi = 0$ and in order for $\rho_c$ to represent the density at the center, $\theta_n(0) = 1$. With these boundary conditions, the equation can be solved, although in most cases only numerically. These are the so-called E-solutions. Given the polytropic index $n$, the density profile is fully determined in terms of the total mass and radius. The mass is given by
\begin{equation}
M = 4 \pi \, \rho_c \, R^3 \, \left.\left(-\frac{1}{\xi} \frac{d \theta_n}{d\xi}\right)\right|_{\xi = \xi_n} \equiv  4 \pi \, \rho_c \, R^3 \, \frac{|\theta_n'(\xi_n)|}{\xi_n} ~,
\end{equation}
where $\xi_n = R/r_n$, so $\theta_n(\xi_n) = 0$. This results in
\begin{equation}
R = \left( \frac{(n + 1) \, K}{\left(4 \pi\right)^{1/n} G} \, \xi_n^{1+1/n} \, |\theta_n'(\xi_n)|^{1 - 1/n}\right)^{- \frac{n}{n - 3}} \, M^{\frac{1 - n}{3 - n}} ~,
\label{eq:MRRpoly}
\end{equation}
which is the well-known polytropic mass--radius relation. Similarly, the central density can be written in terms of the total mass,
\begin{equation}
\rho_c = \left( \frac{(n + 1) \, K}{(4 \pi)^{1/3} \, G} \, \left(\xi_n^2 \, |\theta_n'(\xi_n)|\right)^{2/3}\right)^{\frac{3 n}{n - 3}} \, M^{\frac{2 n}{3 - n}} ~,
\end{equation}
and $\rho(r) = \rho_c \, \theta_n^n(r)$. The escape velocity at the core can be written in terms of the polytropic index and the escape velocity at the surface, $v_e^2(r=0) = \left(1 + \left(\xi_n \, |\theta_n'(\xi_n)| \right)^{-1}\right) \, v_e^2(r=R)$.

These results correspond to zero temperature (full degeneracy). In the case of partial degeneracy, temperature is proportional to the degeneracy parameter and the Fermi energy. For a non-relativistic partially degenerate gas, the Fermi energy is proportional to $\rho^{2/3}$ and pressure to $\rho^{5/3}$, as in the case of a polytrope with index $n = 3/2$. Therefore, $P \sim \rho \, T$, which coincides with the equation of state of an ideal monoatomic gas. In general, for such an equation of state, the temperature profile goes as 
\begin{equation}
T(r) = T_c \, \theta_n(r) \sim \rho(r)^{1/n} \hspace{5mm} \longrightarrow \hspace{5mm} T_c \sim M^{\frac{2}{3 - n}}  \sim \frac{M}{R} ~.
\label{eq:Tpoly}
\end{equation}
Note that this is just the virial theorem for a body of gas in hydrostatic equilibrium. With these tools at hand, along with empirical mass--radius and mass--core temperature relations, we can now describe the relevant properties of different celestial objects. We comment on each type of them below.

\subsection{Planetary bodies}

Planetary bodies are the smallest celestial objects heavy enough so that self-gravity can force them into spherical shape. Among them, those orbiting a star and having cleared the neighborhood around their orbit are called planets, whereas if the last condition is not met they are dwarf planets or satellites. Their composition is classified in three main types: rock made up of silicates (compounds of Mg, Si, O) and iron; ice (water and other nolecules as CH$_4$, NH$_3$, CO, N$_2$ and CO$_2$); and gas (hydrogen and helium).

Both the mechanical and thermal profiles in the interior of planets are fully determined by the equation of state of the material. For spherical bodies of homogeneous composition and at zero temperature, in hydrostatic equilibrium, their radius increases with mass up to a maximum value~\cite{Kothari:1938, Demarcus:1958, Hamada:1961, Zapolsky:1969}, which depends on the planet composition.  Below this critical mass, $\sim \textrm{few}~M_J \sim 10^3~M_\oplus \sim \textrm{few}~10^{-3}~M_\odot$, where $M_J$ is Jupiter's mass,\footnote{Note that a significant H/He envelope would result in a larger radius~\cite{Fortney:2006jm, Seager:2007ix}. Another effect that could enlarge the radius of a planet is irradiation due to the proximity to the parent star~\cite{Baraffe:2003, Batygin:2013}, which slows down contraction. These effects could modify the estimate of the maximum radius at zero temperature. Moreover, for large masses, the hydrogen in the core can fuse and thermal effects become important, so the zero-temperature estimate does not apply. Indeed, young, and still hot, giant planets can be larger than the quoted maximum size. } Coulomb forces balance gravity and the density varies little and roughly corresponds to the value at zero pressure, so that $M \sim R^{1/3}$ (or $n \sim 0$).\footnote{Solid and liquid materials are not infinitely incompressible, so $n > 0$ and thus, the exponent in the mass--radius scaling relation is smaller than 1/3~\cite{Valencia:2005nh, Valencia:2007dt}.} Around this critical mass, pressure is large enough to ionize the material and electron degeneracy pressure starts playing a role. Thus, for $M > \textrm{few}~M_J$, the mass--radius relation tends to $R \sim M^{-1/3}$ (or $n = 3/2$), as will be described below.

The minimum mass required for a planetary body to achieve a nearly spherical shape in hydrostatic equilibrium is estimated to be $\sim 3 \times 10^{-5}~M_\oplus\sim 10^{-10}~M_\odot$~\cite{Tancredi:2009, Lineweaver:2010gg}, which roughly corresponds to the smallest round satellites and dwarf planets in the solar system~\cite{Thomas:2010, Park:2016, Vernazza:2020, Hanus:2019}. Although there exist spherical satellites as massive as $ 3 \times 10^{-4}~M_\oplus$~\cite{Thomas:2010} which are not in hydrostatic equilibrium, in this section we consider the range of masses that represents planetary bodies to be $(3 \times 10^{-5}~M_\oplus \simeq) \, 10^{-10}~M_\odot \lesssim M \lesssim 13~M_J \, (\simeq 0.012~M_\odot)$. 

Regarding the structural properties of planets, although the polytropic model was considered as a first attempt to describe them, that equation of state does not incorporate the approximate incompressibility of solids and liquids at low pressures. To account for it, many different equations of state have been considered~\cite{Murnaghan:1944, Birch:1947, Salpeter:1967, Vinet:1987, Vinet:1989, Poirier:2000, Roy:2005, Roy:2006, Seager:2007ix, Swift:2010nj, Weppner:2015, Mazevet:2019}. The resulting shape of the mass--radius relation, in the case of a homogeneous composition, is very similar for different materials, all cases presenting a maximum radius, with planets with heavier elements resulting in smaller radii~\cite{Fortney:2006jm, Seager:2007ix}. In fact, this can be used to infer the bulk composition of planets. But not only does the size of planets depend on mass, but their composition also does. As a consequence of low-mass cores not being able to efficiently accrete large amounts of gas during the early formation stages, low-mass planets are composed mainly of heavy elements and high-mass planets of gas (see, e.g., Refs.~\cite{Mordasini:2009pp, Mordasini:2012xf, Mordasini:2012hp}). We account for this in a simplified way, by using a variable composition along the following empirical piece-wise mass--radius relation~\cite{Chen:2017} (see also Refs.~\cite{Sotin:2007, Valencia:2007dt, Weiss:2013zma, Hatzes:2015, Zeng:2016, Wolfgang:2016, Bashi:2017}),
\begin{equation}
\left(\frac{R}{R_\oplus}\right) = 
\begin{cases}
\, 1.0 \, \left( \frac{M}{M_\oplus} \right)^{0.28} \hspace{1.15cm} ; \hspace{6mm} 10^{-10} \, M_\odot < M \leq 2 \, M_\oplus \hspace{8.5mm} \textrm{(rocky planets and satellites)} \\[2ex]
\, 0.81 \, \left( \frac{M}{M_\oplus} \right)^{0.59}  \hspace{1cm} ; \hspace{13mm} 2 \, M_\oplus < M \leq 130 \, M_\oplus \hspace{5mm} \textrm{(icy and small gaseous planets)} \\[2ex]
\, 20.6 \, \left( \frac{M}{M_\oplus} \right)^{-0.076} \hspace{7mm} ; \hspace{8mm}  130~M_\oplus < M \leq 13 \, M_J \hspace{8mm} \textrm{(gas giants)}
\end{cases} ~,
\end{equation}
and we have extrapolated the fit down to $M = 10^{-10}~M_\odot$.\footnote{This relation approximately applies to Earth-like satellites and dwarf planets. Note that icy satellites, which are a large fraction of the satellites in our solar system, are less dense and cooler. In any case, these bodies are too small to be of much interest in the context of the capture of DM and the estimate of the DM evaporation mass. We keep them for completeness, but we only consider rock structures.} We only consider the fit in Ref.~\cite{Chen:2017} up to the icy--gas giants limiting mass, $M = 130~M_\oplus$. In the gas giants mass range we ensure continuity with the brown dwarfs regime (see next subsection) with a power-law relation. For brown dwarfs, we do not use the results of Ref.~\cite{Chen:2017} as only a few objects are included, so the fit is likely dominated by giant planets. The parameterization we use results in slightly smaller ($\lesssim 10\%$) radii, which implies smaller DM evaporation masses.

These three regimes correspond to rocky planets (Earth-like planets, but also satellites and dwarf planets), ice and small gas giant planets (Neptune- and Saturn-like) and gas giant planets (Jupiter-like and super-Jupiters). The first transition is expected to be caused by the accretion of substantial volatile gas envelopes~\cite{Weiss:2014}, but also by the range of possible compositions of super-Earths, which can have substantial amounts of water and hence, can have a larger radii. Note that the purely statistical value for this transition obtained with the fit is slightly lower than the theoretically expected minimum mass for planets to retain a gaseous atmosphere by gravitational instability, $\sim 10~M_\oplus$~\cite{Mizuno:1980, Stevenson:1982}. Additionally, planets more massive than $\sim (30-60)~M_\oplus$ are likely to be predominantly composed of gas (hydrogen and helium)~\cite{Laughlin:2015}, which represents the transition between icy and small gaseous giant planets, although this is not visible in the mass--radius relation above. The second transition is likely a consequence of gravitational self-compression as mass increases, such that the growth of planets with mass stops. It is also a consequence of a possible bias in observations, as transiting exoplanets are easier to observe if they are close to their parent star, where irradiation effects are important and could result in larger volumes~\cite{Baraffe:2003, Batygin:2013}.

For rocky planets, the inner temperature is too low to have a significant impact on their size~\cite{Valencia:2005nh, Valencia:2006fx, Fortney:2006jm, Seager:2007ix}. Nevertheless, it is high enough to (totally or partially) melt the interior materials, causing their separation according to density. This results in differentiated planets, with denser material lying beneath less dense material, as is the case, with different degrees, of all planets in our solar system. And the radius of realistic planets with mixed composition lies between that of homogeneous planets composed of the denser and less dense materials~\cite{Fortney:2006jm, Seager:2007ix}. 

Unlike for the Earth, the interior of Earth-like planets is poorly known. Nevertheless, by solving planetary structures with realistic equations of state across the mass range $(0.1 - 30)~M_\oplus$ for two-layer rocky planets, a core radius fraction $\textrm{CRF} = \sqrt{1/3}$ (appropriate for the Earth) is found to represent a reasonable assumption. In this way, we consider as a default model for rocky planets, the following two-layer profile~\cite{Zeng:2017},
\begin{equation}
\rho(r) =
\begin{cases}
\, \frac{1}{\textrm{CRF}} \, \bar{\rho}	\hspace{1.5cm} ; \hspace{1.4cm} 0 \leq r \leq (\textrm{CRF}) \, R \\[2ex]
\, \frac{2}{3} \frac{R}{r} \, \bar{\rho} \hspace{1.65cm} ; \hspace{2mm} (\textrm{CRF}) \, R < r \leq R
\end{cases}
\hspace{5mm} ; \hspace{2mm} \textrm{(rocky planets and satellites)} ~,
\end{equation}
where $\bar{\rho} = 3 \, M / (4 \, \pi \, R^3)$ is the average density of the planet. This profile results in a core mass fraction \linebreak $\textrm{CMF} = \textrm{CRF}^2 = 1/3$. Regarding the constituents, we consider a mass composition with a 2:1 ratio of silicates (MgSiO$_3$) to iron~\cite{Sohl:2015}, with an iron core and a rocky mantle, which is representative of the Earth. Note, however, that the composition can be substantially different, including iron planets and ice satellites. This has little impact on the DM evaporation mass, though. Similarly to other quantities, the DM evaporation mass depends only logarithmically (except at mass-matching values for $v_e < v_d$) on the atomic number of the target particles. We have checked some extreme (unrealistic) composition models and the DM evaporation mass gets modified by $\lesssim 10\%$. Being beyond the scope of this work, we do not pursue a more detailed account of variations on planetary composition, but realistic uncertainties on the DM evaporation mass caused by this are expected to be small.

As the energy transport in the interior of planets is likely dominated by convection within layers~\cite{Sotin:2007}, the temperature profile is quasi-adiabatic within layers. Similar to the simplified density profile, we consider a two-layer constant-temperature profile, with the temperature scaling as $T(r) \propto \rho(r)$ (i.e., assuming a Gr\"uneisen parameter $\gamma = 1$). Estimating the interior temperatures is a rather non-trivial issue, even for the Earth, and depends on the poorly known form of the equation of state at high pressures and its temperature dependence. In this work, we consider a scaling of the core temperature with the mass of Earth-like and dwarf planets and satellites as
\begin{equation}
T_c = 4500~\textrm{K} \, \left(\frac{M}{M_\oplus}\right)^{0.3}  \hspace{1cm} ; \hspace{5mm}  10^{-10}~M_\odot \leq M \leq 2~M_\oplus ~,
\label{eq:Tcrocky}
\end{equation} 
which slightly underestimates the expectations for planets and satellites for $M \gtrsim 10^{-8}~M_\odot$ and slightly overestimates them for smaller masses~\cite{Schubert:2004, Rivoldini:2011, Aitta:2012, Knibbe:2018, Garcia:2019, Li:2020}. For instance, it underestimates Earth's expectations by $\sim 30\%$~\cite{Li:2020}. 

For icy and gaseous giant planets, we consider the density and temperature profiles corresponding to a polytrope of index $n = 1$, although density profiles get steeper with mass as a consequence of the increasing electron degeneracy. This approximation is in reasonable agreement with expectations for continuous radial density profiles of planets like Neptune or Uranus~\cite{Helled:2010} and Saturn or Jupiter~\cite{Marley:2014, Stevenson:2020}. Discontinuous density profiles have also been derived, with a core extending up to 70\% of the planet radius~\cite{Marley:1995, Podolak:2000, Fortney:2010, Nettelmann:2013}; in this case with an equivalent polytropic index $n \sim 0.6 - 1$~\cite{Guillot:2015}. These models cannot be distinguished, however, as they both fit well the measured gravitational potential. 

As for the mass scaling of the core temperature, we use estimates for gas giant planets in the solar system and from Ref.~\cite{Chabrier:2000iu} (for solar metallicity at 5~Gyr) to perform a fit. This facilitates continuity at $M = 13~M_J$ with the relation for brown dwarfs (see below). Note, however, that a significant range of temperatures is possible, depending on a number of factors (atmospheric conditions, composition, age, proximity to the parent star, etc.). The following approximation is intended to describe relatively cool bodies in this mass range,
\begin{equation}
\log \left(\frac{T_c}{10^4~\textrm{K}}\right) = 0.28 + 0.56 \, \log{\left(\frac{M}{M_J}\right)} + 0.22 \, \log^2{\left(\frac{M}{M_J}\right)} + 0.035 \, \log^3{\left(\frac{M}{M_J}\right)} \hspace{5mm} ; \hspace{5mm}  2~M_\oplus < M \leq 13~M_J ~,
\label{eq:TcGP}	
\end{equation} 
and reproduces reasonable well the results in Refs.~\cite{Sotin:2007, Mordasini:2012xf, Nettelmann:2012, Militzer:2013, Miguel:2016, Miguel:2018, Mazevet:2019a} for ice and gas giant planets, in particular, for the ones in our solar system. The increase of the slope with mass is mainly due to the phase transition from molecular to metallic (liquid) hydrogen at around $10^{11}$~Pa and $T \sim (10^3 - 10^4)$~K, when the average energy of electrons becomes higher than the ionization potential of hydrogen (see, e.g., Ref.~\cite{Guillot:2015} and references therein). This results in a faster increase of temperature with pressure, and thus, with mass. Note also that super-Earths are usually warmer (and denser), so Eq.~(\ref{eq:TcGP}) also slightly underestimates the core temperature of the most massive rocky planets~\cite{Valencia:2005nh, Sotin:2007}. This results in a slight underestimation of the DM evaporation mass, as we discuss below.

Concerning the element composition, different materials would likely result in different radius for a given mass. In general, modeling the overall structure and composition is fraught with degeneracies, which we cannot carefully account for in our simplified treatment. Here, we consider a continuous transition from rocky/icy planets to gaseous giant planets based on pebble accretion~\cite{Johansen:2009, Ormel:2010, Bromley:2010, Lambrechts:2012, Morbidelli:2012}, which we parameterize by performing a fit to the results in Ref.~\cite{Lambrechts:2014} corresponding to planet formation at 10 au (about the average distance of Saturn to the Sun). In this way, the mass fraction in elements heavier than hydrogen or helium is given by
\begin{equation}
\log Z = 
\begin{cases}
\,	- 0.015 - 0.042 \, \log\left(\frac{M}{M_\oplus}\right) + 0.027 \, \log^2\left(\frac{M}{M_\oplus}\right) - 0.020 \, \log^3\left(\frac{M}{M_\oplus}\right)  \hspace{5mm} ; \hspace{9mm}  2~M_\oplus \leq M \leq 33.7~M_\oplus \\[2ex]
\, 1.44 - \log\left(\frac{M}{M_\oplus}\right) \hspace{7.85cm} ; \hspace{5mm}  33.7~M_\oplus < M \leq 13~M_J	
\end{cases} ~.
\label{eq:composition}
\end{equation}
The hydrogen plus helium mass fraction is given by $X + Y = 1 - Z$, and we assume solar relative composition (i.e., X/Y = 3). For $M > 2~M_\oplus$, as a benchmark for heavy elements, we consider water. Note that the so-called pebble isolation mass ($M_{\rm PIM} \simeq 34~M_\oplus$ in this case) is smaller the closer the orbit is, which results in a larger fraction of heavy elements for $M \lesssim M_{\rm PIM}$ and in a smaller fraction for $M \gtrsim M_{\rm PIM}$~\cite{Lambrechts:2014}.

\subsection{Brown dwarfs}

Now we consider more massive celestial bodies up to masses for which hydrogen fusion takes place efficiently, which defines stars. As mentioned above, at zero temperature there is a critical mass that sets the maximum size of planets and the transition to brown dwarfs, which is caused by a critical pressure, $\sim (10 - 10^{3})$~GPa (smaller values for lighter elements), above which chemical bonds of the material get broken. Thus, after this pressure is reached, the core of a planet shrinks and its density grows. At this point, ionized electrons start becoming partially degenerate and their pressure increasingly significant. As mass grows, the description of planetary bodies in terms of polytropes experiences a transition from an index $n \gtrsim 0$ (constant density) to $n = 3/2$ (non-relativistic degenerate gas). In the limit of a non-relativistic fully degenerate gas, from Eq.~(\ref{eq:MRRpoly}) we would expect $R \sim M^{-1/3}$. However, Coulomb pressure by ions (which tends to set a constant density with a fixed inter-particle distance scale, $R \sim M^{1/3}$) almost cancels the electron degeneracy pressure, which results in $R \sim M^{-1/8}$ (or $n = 11/9$ for a polytrope)~\cite{Chabrier:2000iu}. The accurate description of these effects is governed by the equation of state around these pressures, which requires accounting for strongly correlated, polarizable, partially degenerate quantum and classical plasmas, in a medium where partial ionization by pressure becomes important. Moreover, the equation of state is not the only factor affecting this relation, which also changes with time as the object cools down, resulting in a smaller radius for older objects and with a mass--radius relation which is even flatter for younger bodies~\cite{Burrows:2011pd}. Thus, in realistic models, the transition between planets and brown dwarfs takes place at higher masses than the prediction at zero temperature due to different effects. 

The transition between giant planets and brown dwarfs is a matter of a long-standing debate~\cite{Stevenson:1991eq, Burrows:1992fg, Chabrier:2000iu, Burrows:2001rb, Baraffe:2008wp, Molliere:2012pb}. In general, it is believed that planets are richer in heavy elements than their parent stars, whereas brown dwarfs share a similar composition, which would hint at a distinction based on the formation mechanism~\cite{Stevenson:1991eq}. Nevertheless, the current distinction is set in terms of mass limits based on nuclear fusion processes. For $M \gtrsim 13~M_J \simeq 0.0124 \, M_\odot$ (for solar metallicity)~\cite{Grossman:1973, Saumon:2008fh, Spiegel:2010ju}, deuterium can start to fuse ($T_c \gtrsim 5 \times 10^5$~K), whereas effective hydrogen burning does not take place for masses below $\sim 0.07 \, M_\odot \simeq 73~M_J$ (which defines stars, $T_c \gtrsim 3 \times 10^6$~K)~\cite{Kumar:1963, Grossman:1974}. Within this range of masses, celestial bodies are usually defined as brown dwarfs.

The empirical mass--radius relation for brown dwarfs can be approximately given by a fit to the results of Ref.~\cite{Chabrier:2000iu} (for solar metallicity and an age of 5~Gyr),\footnote{Gas giants continue to cool down and contract after $\sim$1~Myr, whereas brown dwarfs heat up before $\sim$1~Gyr, but when electron degeneracy dominates, they cool down and contract~\cite{Burrows:1997ka, Chabrier:2000iu, Burrows:2001rb}. In any case, we consider relations relevant for most part of the life of celestial objects, corresponding to late evolutionary stages.}
\begin{equation}
\log\left(\frac{R}{R_J}\right) = 0.063 + 0.0036 \, \log\left( \frac{M}{M_J} \right) - 0.055 \, \log\left( \frac{M}{M_J} \right)^2 \hspace{1cm} ; \hspace{5mm} 13~M_J < M \leq 0.07 \, M_\odot ~. 	
\label{eq:MRRBD}
\end{equation}

As already mentioned, brown dwarfs are well described by non-relativistic partially degenerate polytropes ($n \simeq 3/2$), which are fully determined in terms of their mass and radius. Thus, the mass--radius relation reduces the number of free parameters needed to determine the shape of the density and temperature profiles to one, say the mass. In the case of the density, the constraint on the volume integral (determined by the mass), completely fixes its central value. Thus, our default model follows a polytropic density profile with the temperature profile given by $T(r) \sim \rho(r)^{2/3}$, with the mass--core temperature relation obtained from a fit to the different brown dwarfs in Ref.~\cite{Chabrier:2000iu} (for solar metallicity and an age of 5~Gyr),

\begin{equation}
\log\left(\frac{T_c}{{\rm K}}\right) = 0.34 + 9.36 \, \log\left(\frac{M}{M_J}\right) - 6.20 \, \log^2\left(\frac{M}{M_J}\right) + 1.56 \, \log^3\left(\frac{M}{M_J}\right) \hspace{1cm} ; \hspace{5mm} 13~M_J < M \leq 0.07~M_\odot ~.
\label{eq:TCBD}
\end{equation}
These results are in agreement with others in the literature (see, e.g., Refs.~\cite{Burrows:1997ka, Burrows:2001rb}) and  were obtained with the equation of state for hydrogen and helium from Ref.~\cite{Saumon:1995}, which covers the relevant range of densities and temperatures (see also Refs.~\cite{Nettelmann:2012, Militzer:2013, Becker:2014, Miguel:2016, Miguel:2018, Mazevet:2019a} for recent calculations with ab initio equations of state and Ref.~\cite{Mazevet:2020} for a comparison of them). Note that updated equations of state predict not only higher central temperatures, but also larger radii, and different atmospheric boundary conditions could also result in higher central temperatures~\cite{Becker:2014}. On another hand, metallicities smaller than solar values result in slightly smaller radii and higher core temperatures, although these variations only amount to a few percent (see, e.g., Ref.~\cite{Chabrier:2000iu}) and have a small impact on the calculation of the DM  evaporation mass. At late evolutionary stages, brown dwarfs cool down. We consider $\sim 5$~Gyr as a representative value, but the core temperature of the oldest ($\gtrsim 10$~Gyr) and most massive ($M \gtrsim 40~M_J$) brown dwarfs is expected to be smaller by $\lesssim 20\%$ (see, e.g., Refs.~\cite{Chabrier:2000iu, Burrows:2001rb}). Note, however, that accounting for the formation of dust in the equation of state results in slightly slowing down the cooling process for the most massive brown dwarfs~\cite{Chabrier:2000ns}, so the actual difference in core temperature between 5~Gyr and 10~Gyr is expected to be even smaller.

Finally, for the composition of brown dwarfs, we take 98\% hydrogen and helium (with solar 3:1 proportion) and 2\% heavy elements (water), $X + Y = 0.98$ and $Z = 0.02$~\cite{Chabrier:2000iu, Burrows:2001rb}.

\subsection{Main-sequence stars}

For masses around $\sim 0.07~M_\odot$, celestial bodies contract as they radiate from the surface and the core heats up. This contraction eventually stops either by the appearance of electron degeneracy or by thermal pressure from hydrogen burning. The most dominant process is mainly determined by the attained temperatures. For low enough temperatures, electrons become partially degenerate and hydrogen cannot fuse. Given the weak dependence of pressure on temperature, further radiation from the surface does not result in contraction, but in the cooling of the core. These objects are called brown dwarfs and have been discussed above. On the other hand, if central temperatures are high enough for hydrogen to ignite before electrons become degenerate, energy losses from the surface are balanced by the thermal energy from hydrogen burning and the object inflates. These celestial bodies are called stars. The transition mass between these two types of objects is $\sim 0.07~M_\odot$~\cite{Kumar:1963, Grossman:1974, Burrows:1992fg}. Therefore, thermonuclear reactions involving hydrogen create a thermal pressure that can sustain the gravitational pull of the gas and make stars bigger. This results in the non-degeneracy of the gas, which approximately behaves as a classical ideal gas, $P \propto \rho \, T$. 

In this section we only consider stars during main-sequence. This is the evolutionary phase since ignition after gravitational collapse until the hydrogen core is consumed and converted into helium. This stage is the longest one and represents about 90\% of the stars life, when they can be approximately considered to be in hydrostatic equilibrium, with properties varying relatively little.

For masses above the hydrogen-burning limit, $M \gtrsim 0.07~M_\odot$, electron degeneracy decreases quickly with mass, but up to $M \sim 0.4~M_\odot$, it still impacts star evolution and the inner structure can be described in terms of a polytrope of index $n = 3/2$, as it happens for brown dwarfs. For $M \lesssim 0.2~M_\odot$, stars are fully convective and their internal structure strongly depends on the boundary conditions. As in the case of brown dwarfs, convection deep into the optically thin layers is favored by the formation of molecular hydrogen in the outer parts. The presence of molecules enhances collision-induced absorption, so radiative opacity increases, which in turn, reduces the radiative transport efficiency. Additionally, the adiabatic gradient is reduced, which favors convection~\cite{Kumar:1963}. For slightly heavier stars, convection penetrates less efficiently their inner parts and a radiative core forms. The transition from a convective star to the formation of a significant radiative core occurs at $M \sim 0.4~M_\odot$. However, hydrogen burning is not very efficient up to $M \sim 0.7~M_\odot$ and stars can be approximately described by a polytrope of index $n = 2$ in that mass range (i.e., $0.4~M_\odot \lesssim M \lesssim 0.7~M_\odot$). For $0.7~M_\odot \lesssim M \lesssim 1.5~M_\odot$, the decrease in the central abundance of hydrogen results in an increase of the molecular weight, which heats up further and inflates the star more efficiently. For $M \gtrsim 1.5~M_\odot$, stars reach high enough temperatures ($T_c \gtrsim 2 \times 10^7$~K) so that the CNO burning cycle ignites in most of their core. The extreme temperature dependence of these reactions results in convective instability and the formation of a convective core, which is larger the heavier the star is. This is also favored by the increasing importance of radiation pressure with mass, which reduces the adiabatic gradient.\footnote{Note that the condition for convection instability is that the temperature gradient is steeper than the adiabatic one. This can be achieved either by reducing the adiabatic gradient (as is the case in the outer layers in low-mass stars) or by increasing the temperature gradient (as happens when the CNO cycle is efficient in high-mass stars).} Thus, for masses $M \gtrsim 0.7~M_\odot$, stars can be approximately described by a polytrope of index $n = 3$.

As a consequence of all the above features, from the hydrogen-burning limiting mass to very massive stars, the radius and core temperature continuously increase with mass. At the very low-mass end, $0.07~M_\odot \lesssim M \lesssim 0.4~M_\odot$, the density of stars decreases with mass. Nevertheless, within the mass range $0.4~M_\odot \lesssim M \lesssim 1.5~M_\odot$, the gas is non-degenerate and the nuclear rate of energy production via hydrogen fusion grows in efficiency, resulting in a warmer ($T_c$ with a stronger dependence on the mass than in less massive stars) and a denser core, so the luminosity grows steeply with mass. For $M \gtrsim 1.5~M_\odot$, radiation pressure reduces the mass dependence of the luminosity, and the very steep temperature dependence of the CNO rate results in hotter and less dense stars the more massive they are.

Although a proper treatment of the evolution equations should include the full system of hydrostatic equilibrium, mass conservation, energy transfer and energy conservation equations (along with the chemical equation that governs the changes in composition), the approximation of the density and temperature profiles in terms of polytropes works reasonable well and it is good enough within our simplistic modeling. In particular, as mentioned above, we consider $n = 3/2$ for $0.07~M_\odot \leq M \leq 0.4~M_\odot$, $n = 2$ for $0.4~M_\odot < M \leq 0.7~M_\odot$ and $n = 3$ for $M > 0.7~M_\odot$~\cite{Feiden:2012ri} to obtain the density and temperature profiles. 

As for the other less massive celestial bodies discussed in previous subsections, empirical mass--radius relations for stars have also been proposed for a long time (see, e.g., Refs.~\cite{Torres:2010, Eker:2015, Moya:2018, Eker:2018} for some recent relations). Here, we consider the following empirical mass--radius relation, inspired by and in agreement with the data and models of Refs.~\cite{Chabrier:2000iu, Eker:2018, Parsons:2018}
\begin{equation}
\log \left(\frac{R}{R_\odot}\right) =
\begin{cases} 
\, 1.17 \, \log \left( \frac{M}{M_\odot}\right) + 0.67 \, \log^2 \left( \frac{M}{M_\odot}\right) + 0.43 \, \log^3 \left( \frac{M}{M_\odot}\right)	\hspace{5mm} ; \hspace{5mm} 0.07~M_\odot < M \leq 1.5~M_\odot \\[2ex]
\, 0.13 + 0.56 \, \log \left( \frac{M}{M_\odot}\right)	\hspace{5.1cm} ; \hspace{6mm} 1.5~M_\odot < M \leq 100~M_\odot
\end{cases}	~.
\label{eq:MRR_stars}
\end{equation}	
For low-mass stars, we perform a fit using the data compiled in Ref.~\cite{Parsons:2018}, which is in good agreement with Refs.~\cite{Chabrier:2000iu, Eker:2018} for $M \leq 1.5~M_\odot$. The (log) linear parameterization in the high-mass range is a simplification of the piece-wise fit in Ref.~\cite{Eker:2018}. 
 
As for any polytropic model, this relation reduces the number of parameters required to describe the internal stellar structure to one and, as done throughout this paper, we use the mass of the star. Note that using the mass--radius relation above, the core density, $\rho_c \sim M/R^3$, always decreases with the stellar mass. Nevertheless, we have mentioned above that in the mass interval $0.4~M_\odot \lesssim M \lesssim 1.5~M_\odot$, detailed stellar models predict the core density to slightly grow with mass. We cannot correct for this trend using the polytrope assumption, but given that it is known to provide an approximately correct overall description, we use it in what follows. This approximation has a negligible impact on the calculation of the DM evaporation mass.

The situation is different for the core temperature, which is not fixed by the mass--radius relation. We obtain a monotonously increasing mass--core temperature relation using the results corresponding to the models for low-mass stars from  Ref.~\cite{Chabrier:2000iu} and for high-mass stars from Ref.~\cite{Kippenhahn:1994wva} (at zero-age main sequence),
\begin{equation}
\log\left(\frac{T_c}{10^4~\textrm{K}}\right) = 
\begin{cases}
3.17 + 0.82 \, \log\left(\frac{M}{M_\odot}\right) - 0.25 \, \log^2\left(\frac{M}{M_\odot}\right) - 2.15 \, \log^3\left(\frac{M}{M_\odot}\right) - 1.60 \, \log^4\left(\frac{M}{M_\odot}\right) \\ 
\hspace{10.5cm} ; \hspace{5mm} 0.07~M_\odot < M \leq 1.5~M_\odot \\
3.24 + 0.30 \, \log\left(\frac{M}{M_\odot}\right) - 0.054 \, \log^2\left(\frac{M}{M_\odot}\right) \hspace{5mm} ; \hspace{5mm} 1.5~M_\odot < M \leq 100~M_\odot 
\end{cases} ~.
\end{equation}
We have also confirmed that these are reasonable parameterizations by comparing them with the evolutionary tracks obtained with the MIST code~\cite{Dotter:2016, Choi:2016}, that in turn uses the stellar evolution code MESA~\cite{Paxton:2010, Paxton:2013, Paxton:2015}. 

Note that for $M \sim (0.2 - 0.4)~M_\odot$, the core temperature is roughly given by $T_c \simeq 1.6 \, \mu_{\rm mol} \, G \, M/(3 \, R)$, where $\mu_{\rm mol}$ is the mean molecular weight of the material (in units of the atomic mass unit), which is $\mu_{\rm mol} \simeq 0.6$ for solar abundances. This is the expected result for a polytrope of index $n = 3/2$, for a monoatomic and fully ionized ideal gas, Eq.~(\ref{eq:Tpoly}). This is a consequence of degeneracy becoming less important with mass, while the star being largely convective. For $0.4~M_\odot \lesssim M \lesssim 1.5~M_\odot$, had we used the mass--radius relation above, assuming the equation of state of an ideal gas would not result in $T_c$ increasing with mass, though. For higher masses, $M \gtrsim 1.5~M_\odot$, radiation pressure grows in importance, which results in a dependence of the core temperature on the mass less steep than at lower masses, as mentioned above.

Furthermore, unlike what happens for less massive celestial objects (planetary bodies and brown dwarfs), stars remain in main sequence for a period of time which is shorter the heavier the star is. It is, thus, relevant to discuss how long this stage lasts, such that their overall properties remain approximately constant, without much change. For the DM capture process to be efficient and the discussion about the DM evaporation mass to be of most interest, the equilibration time $\tau_{\rm eq}$ (actually, $\tau_{\rm eq}/\kappa$) must be shorter than the stars age. Otherwise, the DM evaporation mass is time dependent.

As the total energy release in the $pp$ chain and the CNO cycle is the same, a simple estimate for the lifetime in main sequence is given by
\begin{equation}
t_{\rm life, MS} = t_{\rm life, \odot} \, \left(\frac{M/M_\odot}{L/L_\odot}\right) ~,
\end{equation}
where $t_{\rm life, \odot} = 10$~Gyr and $L_\odot = 3.828 \times 10^{26}$~W are the lifetime and current luminosity of the Sun. To obtain the lifetime of other stars we consider the empirical mass--luminosity relation from Ref.~\cite{Eker:2018}, which we extrapolate, in the low- and high-mass extremes, to cover the entire stellar mass range discussed in this work. Moreover, we simplify the multiple piece-wise form by performing a smooth fit in the mass interval $0.1~M_\odot \lesssim M \leq 100~M_\odot$,
\begin{equation}
	\log \left(\frac{L}{L_\odot}\right) = 4.24 \, \log \left(\frac{M}{M_\odot}\right) + 0.239 \, \log^2 \left(\frac{M}{M_\odot}\right) -0.759 \, \log^3 \left(\frac{M}{M_\odot}\right) + 0.216 \, \log^4 \left(\frac{M}{M_\odot}\right) ~.
\end{equation}
Notice that only stars with masses $M \lesssim 2~M_\odot$ have lifetimes in the main sequence longer than $\sim 1$~Gyr. Very massive stars, $M \gtrsim 10~M_\odot$ leave the main sequence after less than $\sim 10$~Myr.

\subsection{Post-main-sequence stars}

The internal properties of stars during post-main-sequence evolution undergo significant changes in relatively short periods of time, which crucially depend on the stellar mass they had in main sequence and on environmental conditions~\cite{Kippenhahn:1994wva}. Given that evaluating the implications of non-static configurations on the DM evaporation mass is beyond the scope of this paper, we briefly and qualitatively describe some of the stellar properties after stars leave the main sequence in this subsection, and leave the description of compact stellar remnants to the next subsections. 

The main-sequence phase ends when hydrogen is exhausted in the stellar core. At this point, an approximately isothermal helium core develops and hydrogen burns in a shell surrounding it~\cite{Iben:1984}. This results in the production of more helium, which increases the mass of the core, bounded by the Sch\"onberg-Chandrasekhar limit~\cite{Schonberg:1942}, and reduces its size. While the core temperature slightly increases, the envelope of the star expands and cools, becoming mostly convective, to favor the transport of additional energy. This is the so-called subgiant branch for low- and intermediate-mass stars and lasts for $\sim \textrm{few Gyr}$ for $M \lesssim 1~M_\odot$ and $\sim 10^7$~yr for $M \gtrsim 2~M_\odot$. During this phase, although the core temperature does not change much (it is just slightly higher than the terminal age main-sequence value), the central density can increase by more than an order of magnitude with respect to the main-sequence value~\cite{Iben:1967}.

Eventually, stars enter the red giant branch, delimited by their Hayashi line, and their luminosity increases. Very low-mass stars ($M \lesssim 0.3~M_\odot$), however, are fully convective and do not reach temperatures high enough to burn helium, so they do not become red giants and end up as helium white dwarfs~\cite{Laughlin:1996}, when all hydrogen has been fused to helium. For low-mass stars, $0.3~M_\odot \lesssim M \lesssim 2.3~M_\odot$, the density of the core can get so high that electrons become degenerate and for $M \gtrsim 0.8~M_\odot$, the temperature increases up to a point when helium starts to burn ($T_c \sim 10^8$~K)~\cite{Iben:1984} and the core mass is $\sim 0.5~M_\odot$~\cite{Buzzoni:1983}. At this point, as temperature rises (in a run-away process, as pressure does not depend on temperature), the core turns non-degenerate again, which finally results in a helium flash.\footnote{Helium burning does not start at the center of the core. Large amounts of energy lost in the form of neutrinos during the final stages in the red giant branch produce a temperature inversion, so that helium burning starts in a shell around the helium-rich core~\cite{Thomas:1967, Paczynski:1977, Sweigart:1978}.}  This phase lasts for about $10^9$~yr for $M \sim 1~M_\odot$ and $10^7$~yr for $M \sim 2~M_\odot$. During this phase, the core temperature increases by a factor of a few, and the central density is several orders of magnitude larger than during main sequence, although it decreases much faster towards the edge of the core. On the other hand, the stellar radius can increase by two orders of magnitude for the lightest stars and by a factor of a few for the most massive ones in this interval, and a significant mass loss is experienced. 

The end of this phase for low-mass stars, which results in stellar contraction and core cooling, marks the entry into the horizontal branch stage. This phase lasts for up to $\sim 10^8$~years~\cite{Buzzoni:1983}, while helium burns and a carbon-oxygen core develops, and it is similar, but shorter, than the hydrogen-burning phase that defines main sequence. The position on the horizontal branch depends on the mass lost during the red giant phase. The larger the mass loss, the bluer and slightly smaller the star becomes. In particular, a $\sim 1~M_\odot$ star on this branch is an order of magnitude larger than during main sequence.

After the helium core is depleted, the immediate stellar evolution strongly depends on the mass~\cite{Iben:1983, Herwig:2005, Karakas:2015}. For low masses ($0.3~M_\odot \lesssim M \lesssim 2.3~M_\odot$), stars enter the asymptotic giant branch, which is similar to the red giant branch, but with slightly higher temperatures. The non-degeneracy of the core, however, implies that the increase of pressure leads to expansion and cooling of the region where helium burns. 

In heavier stars, $2.3~M_\odot \lesssim M \lesssim (8-10)~M_\odot$~\cite{Iben:1983, Herwig:2005, Karakas:2015, Cristallo:2015}, helium ignites before the core becomes degenerate, so there is no helium flash. Stars can move away from the red giant branch getting bluer and then becoming redder again (blue loops), and a degenerate carbon-oxygen core develops. The most massive stars in this mass range might not even reach the red giant branch before turning into red supergiants (if they retain their envelope) or getting bluer and becoming Wolf-Rayet stars (if they do not retain their envelope). This phase lasts for a short period of time, the shorter the more massive stars are. All these stars, $M \lesssim (8-10)~M_\odot$, end up their lives as white dwarfs. For more massive stars, $(8 -10)~M_\odot \lesssim M \lesssim 25~M_\odot$, the helium-burning phase is also the red supergiant stage. These stars, though, proceed through a series of nuclear burning phases and end up as core-collapse supernovae, leaving neutron stars as remnants. The remnants of even more massive stars are black holes.

\subsection{White dwarfs}

Once stars with initial masses $M \lesssim (8-10)~M_\odot$ reach the tip of the asymptotic red giant branch, they attain their maximum size and continue losing mass at a fast pace, ejecting whole shells of material (creating planetary nebulae). At this point, stars get hotter and become blue supergiants in a very short period of time, $\sim 10^4$~yr, until only the hot core remains. This results in a narrow final range of masses. However, temperatures are not high enough for nuclear fusion to proceed and, lacking any other energy sources, stars contract. This increase in density finally results in electron degeneracy throughout the star, which creates a new source of pressure that halts gravitational contraction. From then on, they become white dwarfs (see, e.g., Refs.~\cite{Camenzind:2007, Winget:2008iu, Althaus:2010, Kippenhahn:1994wva, Corsico:2019}) and this is the final evolutionary stage for stars with masses $M \lesssim (8-10)~M_\odot$~\cite{Woosley:2015}, which represent the fate of about 97\% of all stars. Due to electron degeneracy, energy losses from the surface cannot be balanced any more by reducing the stellar size, so white dwarfs keep on cooling for the rest of their lives, unless they accrete matter from a nearby star. 

Analogously to the description of polytropes, the mechanical properties of white dwarfs can be decoupled from the thermal ones. Nevertheless, the equation of state is not that of a polytrope. In this case, pressure is generated by a gas of fully degenerate electrons, but with a varying degree of relativistic Fermi momentum. The equation that describes the internal structure of white dwarfs, which is similar to the polytrope model equation, is the Chandrasekhar equation~\cite{Chandrasekhar:1935}, 
\begin{equation}
\frac{\dd^2 \varphi}{\dd \zeta^2} + \frac{2}{\zeta} \, \frac{\dd \varphi}{\dd \zeta} + \left(\varphi^2 - \frac{1}{z_c^2}\right)^{3/2} = 0 ~,
\end{equation}
with boundary conditions $\varphi(0) = 1$ and $\varphi'(0) = 0$, and
\begin{equation}
	\varphi \equiv \frac{z}{z_c}  \hspace{1cm} ; \hspace{1cm} \zeta \equiv 
	\left(\mu_e \, m_u \, m_e \, \sqrt{\frac{4 \, G}{3 \, \pi}} \, z_c\right) \, r  ~. 
\end{equation}
Here $z^2 = \left(p_{\rm F}/m_e\right)^2 + 1$, with $p_{\rm F}$ the Fermi momentum, and $z_c = z(r = 0)$, with $z \in [1, \infty)$. Note that this equation reduces to the Lane-Emden equation for polytropes in the limits of $z \to \infty$ ($n = 3$) and $z \to 1$ ($n = 3/2$).

The density profile, radius and mass of white dwarfs are given by
\begin{eqnarray}
\rho(r) & = & \left(\frac{\mu_e \, m_u \, m_e^3}{3 \, \pi^2}\right) \left(z_c^2 \, \varphi^2 - 1\right)^{3/2} \simeq 1.95 \times 10^6~\textrm{g/cm}^3 \, \left(z_c^2 \, \varphi^2 - 1\right)^{3/2} ~, \\
R & = & \sqrt{\frac{3 \, \pi}{4 \, G}} \, \frac{\zeta_1}{\left(\mu_e \, m_u \, m_e \, z_c\right)} \simeq 0.61 \, \frac{\zeta_1}{z_c} \, R_\oplus ~, \\
M & = & \frac{\sqrt{3 \, \pi}}{2} \, \frac{1}{(\mu_e \, m_u)^2 \, G^{3/2}} \, \left| \zeta_1^2 \,  \varphi'(\zeta_1)\right| \simeq 0.72 \, \left| \zeta_1^2 \, \varphi'(\zeta_1)\right| \, M_\odot ~,  
\end{eqnarray}
where $\mu_e \simeq 2/(1 + X)$ is the mean molecular weight per free electron and $\zeta_1 \equiv \zeta (z = 1)$. The Chandrasekhar solution does not imply a lower bound on the radius ($R \to 0$ for $z_c \to \infty$), which decreases as the central density increases, keeping the total mass at a fixed value, $M_{\rm Ch} \simeq (2/\mu_e)^2 \, 1.456~M_\odot$; this is the so-called Chandrasekhar mass. For $z_c \to 1$, $\zeta_1 \to \infty$ and $\left| \zeta_1^2 \, \varphi'(\zeta_1)\right| \to 0$ and thus, $R \to \infty$ and $M \to 0$. These limiting conditions are modified once several corrections (general relativity, equation of state, Coulomb interactions) are incorporated. The resulting maximum values, which slightly depend on composition, are $M_{\rm max} \sim 1.3~M_\odot$ and $R(M_{\rm max}) \sim 0.02~M_\odot$.

\begin{figure}[t]
	\centering
	\includegraphics[width=0.49\linewidth]{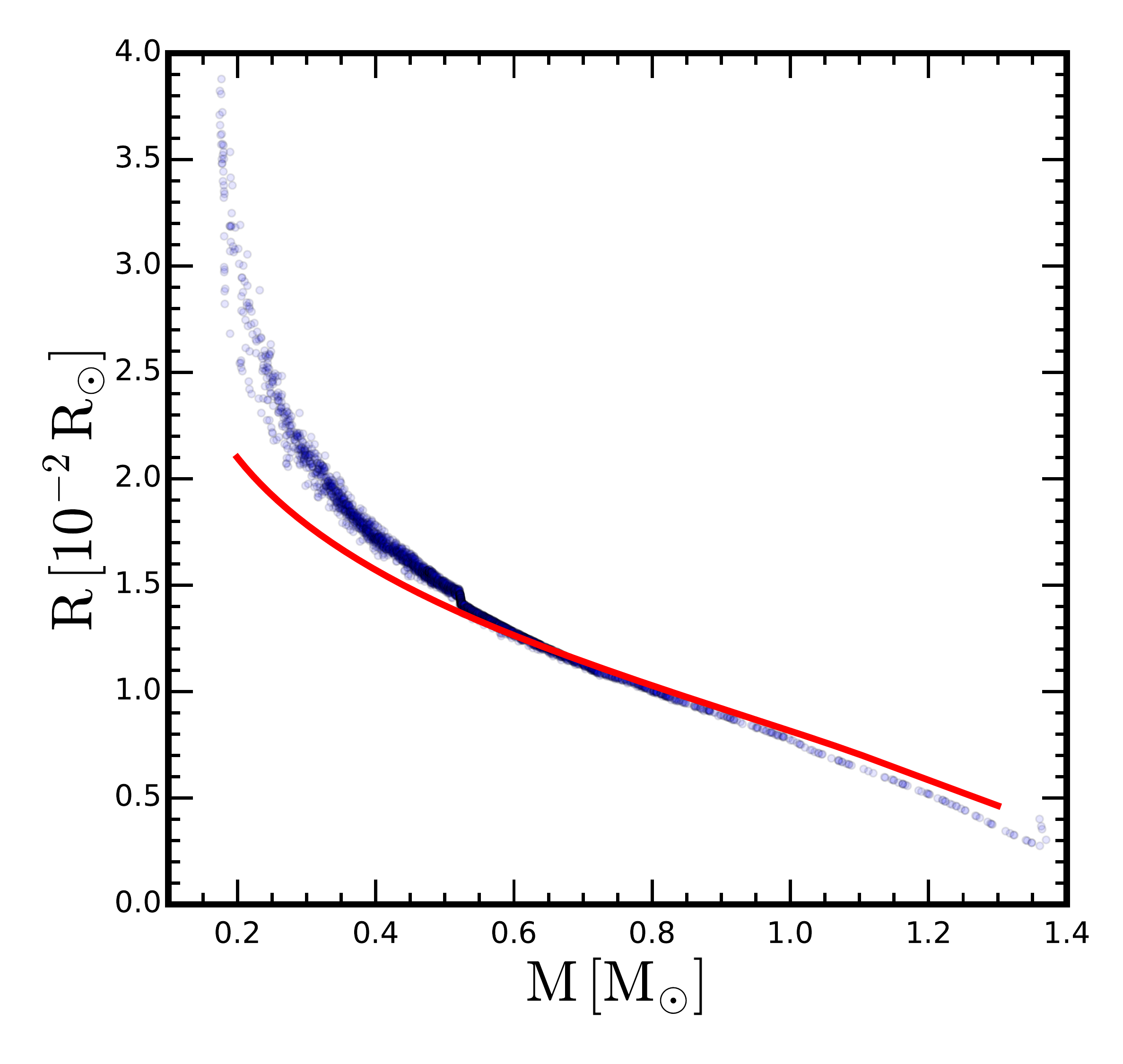}
	\includegraphics[width=0.49\linewidth]{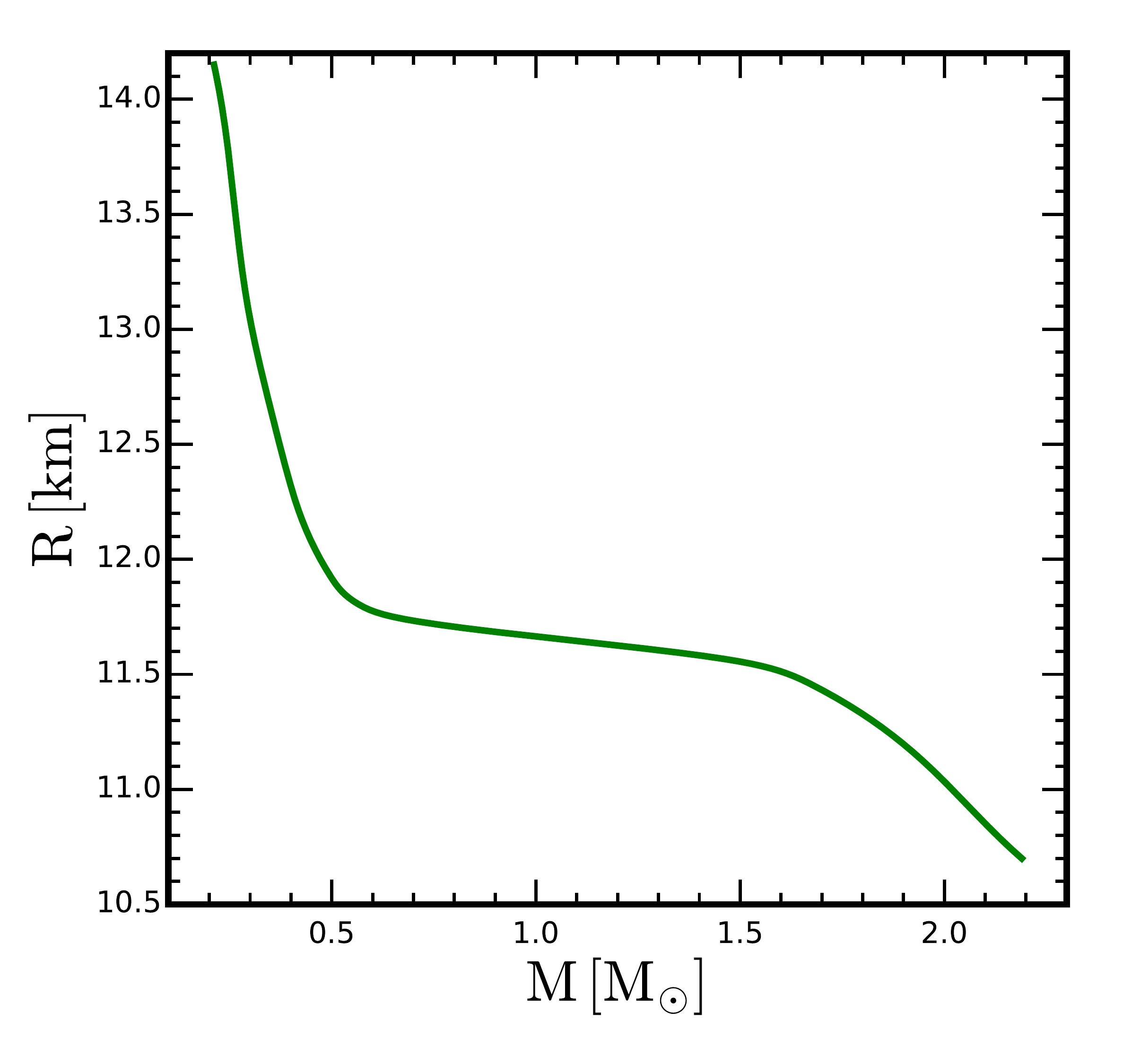}
	\caption{\textbf{\textit{Mass--radius relation}} for white dwarfs (left panel) and neutron stars (right panel). We also show a selection of observed cool white dwarfs with $T_{\rm eff} < 10^4$~K and $\textrm{S/N} \geq 10$~\cite{Kepler:2019}. For neutron stars, we use the equation of state from Ref.~\cite{Potekhin:2013qqa}, based on the nuclear energy-density functional BSk20~\cite{Goriely:2010bm}. }  	
	\label{fig:MRR-compact}
\end{figure}

White dwarfs have been observed with masses $M \sim (0.2 - 1.3)~M_\odot$, with a distribution peaked at $M \simeq 0.6~M_\odot$~\cite{GentileFusillo:2019, Kepler:2019}. Here, we consider this range of masses and solve the Chandrasekhar equation to obtain the mass--radius relation and the density profile. The former is shown in the left panel of Fig.~\ref{fig:MRR-compact}, along with a selection of observed cool white dwarfs with $T_{\rm eff} < 10^4$~K~\cite{Kepler:2019}, which illustrates that Chandrasekhar solution represents a reasonable approximation, and in general, a conservative one for the calculation of the DM evaporation mass.

Most white dwarfs have a carbon-oxygen core surrounded by a thin helium envelope ($Y \lesssim 0.01$), which is in turn surrounded by an even thinner hydrogen envelope ($X \lesssim 10^{-4}$), although low-mass white dwarfs can have a helium core (those stars not massive enough, $M \lesssim 0.3~M_\odot$, to burn helium) and the most massive white dwarfs ($M \gtrsim 1~M_\odot$) can develope an oxygen-neon core. As representative of the core composition, we consider white dwarfs with $Z_{\rm C} = 0.4$ and $Z_{\rm O} = 0.6$ (see, e.g., Refs.~\cite{Straniero:2002pf, Aliotta:2016ove}), although differences on the composition do not affect our results of the DM evaporation mass. 

In order to approximately describe the thermal properties of white dwarfs (at least of relatively cool and evolved ones), we point out that the core constitutes more than 99\% of their mass and that the main contribution to the heat capacity comes from the non-degenerate gas of ions. Degenerate electrons are very efficient in transporting energy outwards, so the core can be approximately described to be isothermal. Finally, energy is radiated away through the non-degenerate envelope, which cools down the white dwarf. A simple description in terms of a two-layer model~\cite{Mestel:1952} provides a very good agreement with more refined predictions of the cooling evolution of white dwarfs~\cite{Iben:1984b, Chabrier:2000ib}. Thus, in this work we consider an isothermal core with a temperature in the range $4 \times 10^5~\textrm{K} \leq T_c \leq 4 \times 10^6~\textrm{K}$, which is appropriate for white dwarfs older than $\sim 3$~Gyr~\cite{Chabrier:2000ib}.

\subsection{Neutron stars}

In even heavier stars than those discussed above, $(8 - 10)~M_\odot \lesssim M \lesssim 25~M_\odot$, the carbon-oxygen core reaches high enough temperatures ($T_c \gtrsim 10^9$~K) to continue nuclear fusion processes before becoming degenerate, resulting in a shell-like structure, which ends up in an iron core for $M \gtrsim 10~M_\odot$. Once nuclear fusion cannot proceed further, stars approach the end of their lives as core-collapsed supernovae explosions (or via electron-capture supernovae in a narrow range of masses). The remnants of these explosions are neutron stars (see, e.g., Refs.~\cite{Haensel:2007, Camenzind:2007, Kippenhahn:1994wva}), which are born very hot ($T \gtrsim 10^{10}~$~K), but cool down very fast by neutrino emission (which lasts for $\lesssim 10^5$~yr), down to $T \sim 10^8$~K after $\sim$100~yr. As density increases, neutron-rich nuclei start releasing free neutrons, which are increasingly degenerate and would eventually be the main source of pressure. Further increase of the density would result in the formation of a degenerate bath of neutrons plus a small admixture of electrons, muons and protons.

The description of the interior of neutron stars is particularly challenging. Unlike main-sequence stars or other post-main-sequence phases, the huge gravitational field of neutron stars requires the use of general relativity. This modifies the Newtonian hydrostatic equilibrium equation, which is replaced by the Tolman-Oppenheimer-Volkoff equation~\cite{Tolman:1939, Oppenheimer:1939}. This just represents a calculational complication, but the most important issue is that the correct equation of state to be used is not yet known. Nevertheless, the determination of relatively large masses for neutron stars from gravitational wave observations of merging systems of binary neutron stars and of a black hole and a neutron star, favor stiff equations of state, which predict relatively large maximum masses, $\sim (2 - 3)~M_\odot$~\cite{Alsing:2017bbc, Margalit:2017dij, Shibata:2017xdx, Ruiz:2017due, Rezzolla:2017aly, Shibata:2019ctb, Shao:2019ioq, Li:2021crp, Nathanail:2021tay}. In this work, we consider the unified equation of state from Ref.~\cite{Potekhin:2013qqa}, based on the nuclear energy-density functional BSk20~\cite{Goriely:2010bm}, which covers the mass range $ 0.09~M_\odot < M < 2.3~M_\odot$. The resulting mass--radius relation is shown in the right panel of Fig.~\ref{fig:MRR-compact}. Notice that differences with respect to more recent equations of state~\cite{Pearson:2018tkr} have a negligible impact on the results presented in this work.

For old neutron stars ($t_{\rm NS} \gtrsim 10^5$~yr), the main cooling mechanism is electromagnetic cooling. At this evolutionary stage the internal temperature is expected to be similar to the surface one and, in idealized scenarios, temperatures as low as $\sim 10^3$~K after $\sim 10^7$~yr, and even lower for older neutron stars, are expected~\cite{Yakovlev:2004iq}. Nevertheless, for theroretical models to be consistent with observations, a heating mechanism is required. Different possibilities have been suggested, which could result in temperatures as high as $10^6$~K~\cite{Gonzalez:2010}. Therefore, we consider an isothermal profile within the interval $10^5~\textrm{K} \leq T_c \leq 10^6~\textrm{K}$, as representative of relatively old neutron stars, $t_{\rm NS} > 10^7$~yr. Neutron stars younger than $\sim 10^6$~yr are expected to have typical central temperatures of $\sim 10^8$~K~\cite{Page:2004fy}.

\section{DM evaporation mass in celestial bodies}
\label{sec:results}

After the overall description of celestial bodies in the previous section, we now compute the minimum mass DM particles must have, such that evaporation from the capturing object is not efficient. The calculation of all the elements required for the computation of the DM evaporation mass follows Ref.~\cite{Garani:2017jcj}, including the correction to the capture rate from Ref.~\cite{Busoni:2017mhe}, and Ref.~\cite{Garani:2018kkd} for neutron stars, as described in Section~\ref{sec:basics}. We first evaluate the DM evaporation mass for the geometric cross section, $\sum_i N_i \, \sigma_i^{\rm geom} = \pi \, R^2$, which results in a capture rate close to its maximum value (the saturation value), for planetary objects, brown dwarfs and main-sequence stars. This cross section is different for each object and depends on the DM mass. Next, we compute the DM evaporation mass for all those objects and within a wide range of cross sections, in the thin and thick regimes. For post-main-sequence stellar phases we qualitatively discuss the evolution of the DM evaporation mass, and for white dwarfs and neutron stars we compute the DM evaporation mass for the geometric cross section. All results are obtained for SI interactions with constant scattering cross section, but we also comment on the DM evaporation mass in the SD case. Moreover, we study its dependence with several factors.

\subsection{DM evaporation mass in planetary bodies, brown dwarfs and main-sequence stars}

\begin{figure}[t]
	\centering
\hspace*{-0.65cm}	\includegraphics[width=1.05\linewidth]{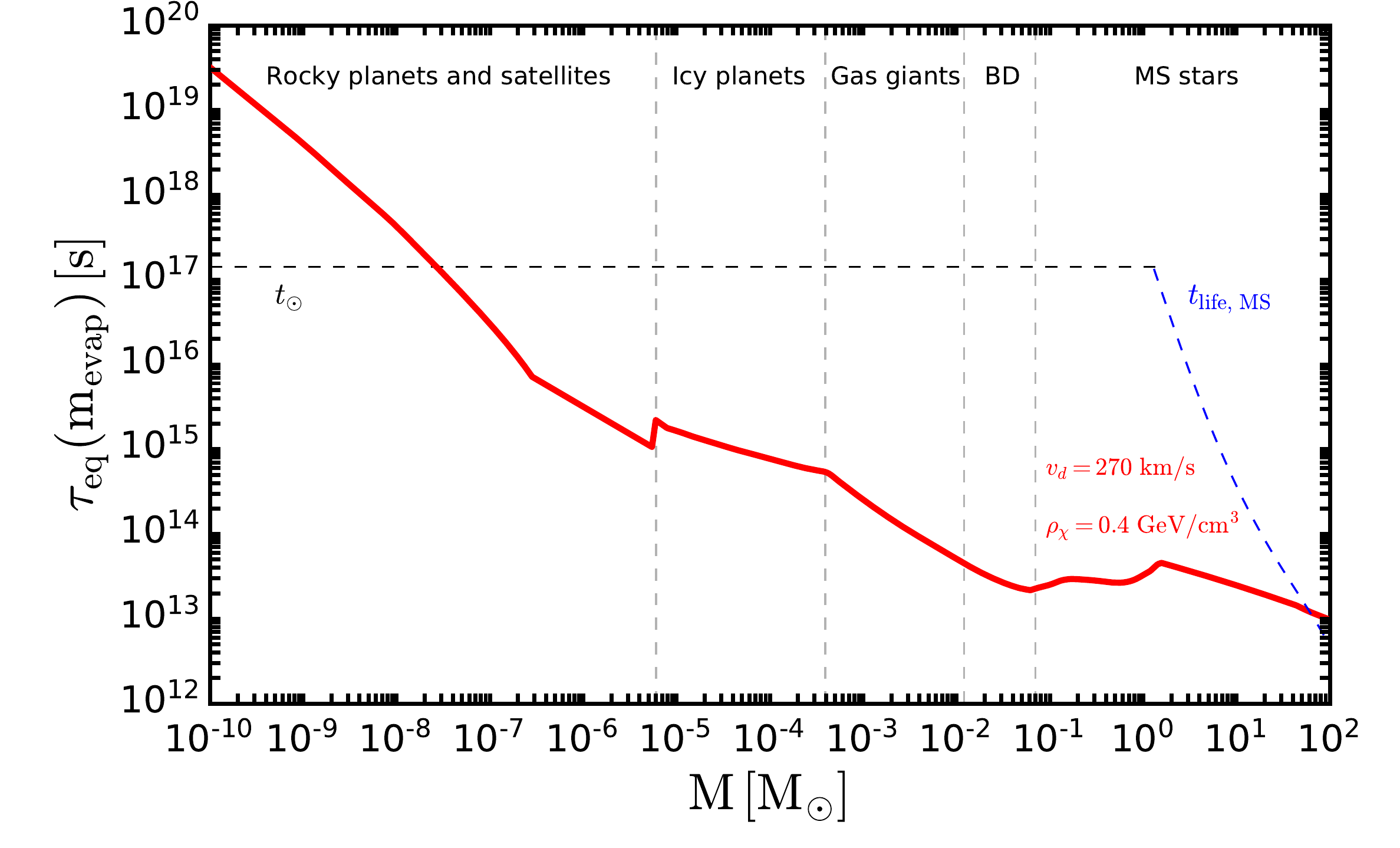}\\
	\caption{\textbf{\textit{DM equilibration time}}, for the DM evaporation mass, as a function of the mass of the capturing object, for planetary bodies, brown dwarfs and main-sequence stars. We take the geometric SI cross section, $\sum_i N_i \, \sigma_i^{\rm geom} = \pi \, R^2$, and $\langle \sigma_A v_{\chi \chi}\rangle = 3 \times 10^{-26}~\textrm{cm}^3/\textrm{s}$. Also depicted are the current solar age, $t_\odot = 4.5$~Gyr (black dashed line) and the stellar lifetime, $t_{\rm life, MS}$, when shorter than $t_\odot$ (blue dashed line). The jump at $M = 2~M_\oplus$ is mainly due to the non-smooth transition in composition and the fact that $v_e < v_d$.}  	
	\label{fig:taueq}
\end{figure}

Before discussing the main results, it is important to evaluate whether equilibrium between DM capture and annihilation is reached for each particular celestial body. Otherwise, the DM evaporation mass is not constant but grows with time, assuming the properties of the capturing object remain approximately unchanged. For very low-mass objects, the equilibration time can be much longer than their age, and even than the age of the Universe. For intermediate-mass and the most massive stars, the equilibration time (for the DM evaporation mass) decreases with stellar mass, although with a much weaker dependence than the lifetime. 

\begin{figure}[t]
	\centering
\hspace*{-0.65cm}	\includegraphics[width=1.05\linewidth]{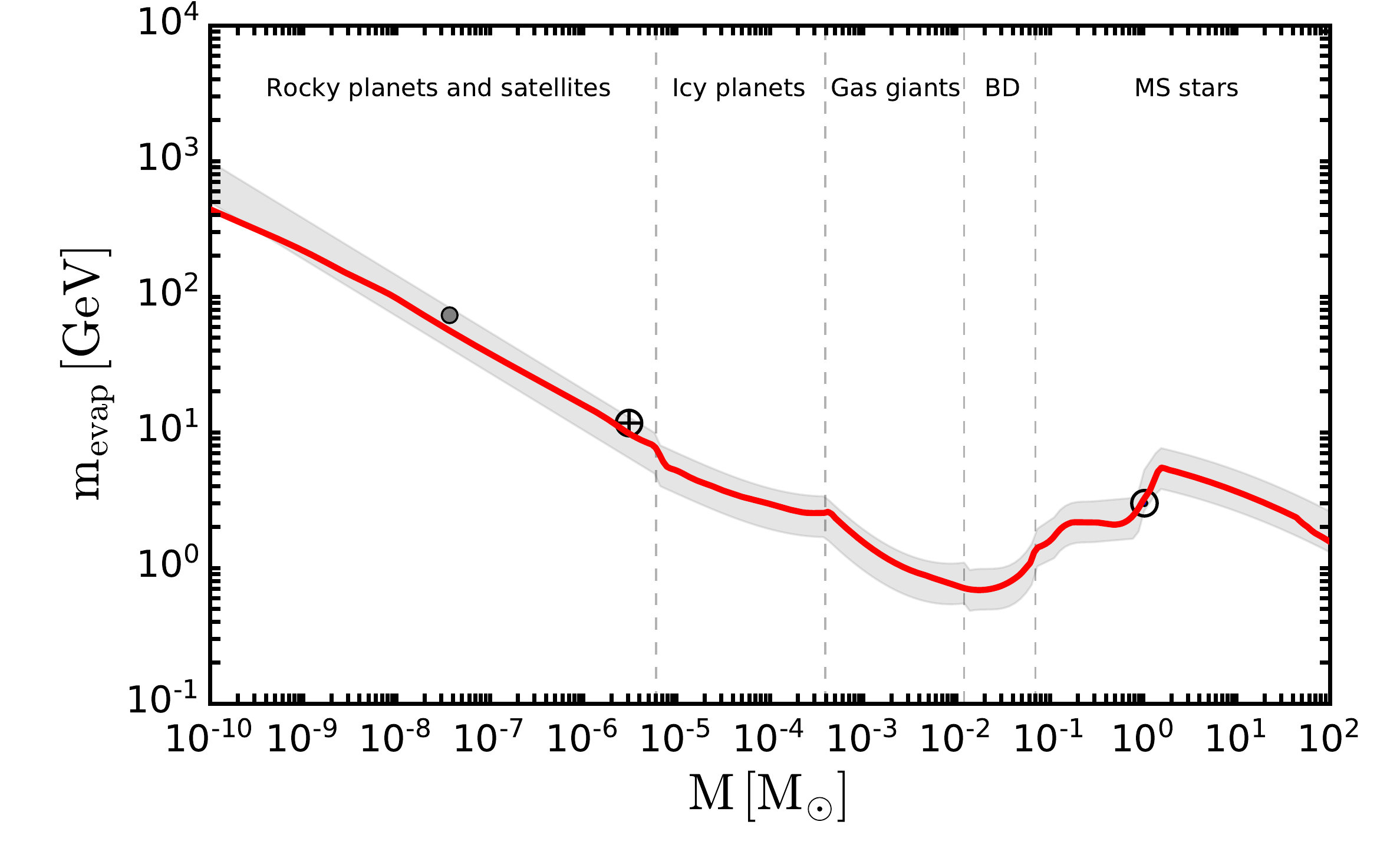}
	\caption{\textbf{\textit{DM evaporation mass}} as a function of the mass of the capturing object, for planetary bodies, brown dwarfs and main-sequence stars. We take the geometric SI scattering cross section, $\sum_i N_i \, \sigma_i^{\rm geom} = \pi \, R^2$, the canonical value of the DM annihilation cross section $\langle \sigma_A v_{\chi \chi} \rangle = 3 \times 10^{-26}~\textrm{cm}^3/\textrm{s}$, and assume a position within the local neighborhood, $\rho_\chi = 0.4~\textrm{GeV}/\textrm{cm}^3$ and $v_d = 270~\textrm{km/s}$, although with other values similar results are obtained. We also indicate the DM evaporation mass using detailed models and data for the Moon, Earth and Sun. The shaded band depicts the range $E_c/T_c = (20 - 40)$.}
	\label{fig:evap_geom}
\end{figure}

In Fig.~\ref{fig:taueq} we compare the equilibration time, $\tau_{\rm eq}$, corresponding to the geometric cross section and to the DM evaporation mass shown in Fig.~\ref{fig:evap_geom}, with other characteristic time scales (solar system current age, $t_\odot \simeq 4.5$~Gyr, and stars lifetime). This comparison is shown as a function of the mass of celestial bodies, which spans the range $10^{-10}~M_\odot \leq M \leq 10^2~M_\odot$, from small satellites to massive stars. The general trend, although it does not apply to the entire mass range, can be understood from the discussion in Section~\ref{sec:basics}: the equilibration time for the DM evaporation mass decreases with the mass of the capturing body. The discontinuity at $M = 2~M_\oplus$ can be understood from our modeling of a non-smooth transition in composition and the fact that $v_e < v_d$, so the $1/\mu_{-}^2$ factor plays a key role (see Eq.~(\ref{eq:eqmevapplanets})). As already mentioned, if the equilibration time is longer than the age of the object, the DM evaporation mass grows with time, as the number of capture DM particles does (as long as the properties of the object do not change). This occurs until equilibration is reached or the system is destroyed or it evolves in a significant manner. 

For the local DM density, the canonical thermal annihilation cross section and the geometric cross section, DM capture and annihilation do not reach equilibrium for the smallest objects, $M \lesssim 3 \times 10^{-8}~M_\odot$, during the solar system age ($t_\odot = 4.5$~Gyr), so the DM evaporation mass would increase with time. Likewise, equilibration does not take place during the lifetime of the most massive stars, $M \gtrsim 60~M_\odot$. Furthermore, planetary systems are usually bounded to stellar objects, so their lifetimes could be linked to stellar lifetimes. Objects in the range $2 \times 10^{-8}~M_\odot \lesssim M \lesssim 60~M_\odot$ are in the regime in which the DM capture and annihilation rates could be in equilibrium, so the DM evaporation mass would remain approximately constant, as long as the properties of the objects can be regarded as constant. This is so for the DM annihilation and scattering cross sections and the DM density and velocity dispersion we have considered in this figure. Larger annihilation cross sections, higher DM densities or lower velocity dispersion, would result in shorter equilibration times. For instance, for the smallest objects or the most massive stars, if close to the galactic center, equilibrium is more likely. Note that capture is already assumed to be close to maximum, so the equilibration time could only increase by considering a smaller scattering cross section and remains almost unchanged for larger cross sections.

In Fig.~\ref{fig:evap_geom}, we show the DM evaporation mass as a function of the mass of the capturing object, for the geometric SI cross section and assuming $t = \textrm{min}\{t_\odot, \tau_{\rm life, MS}\}$. We see that it decreases from values $m_{\rm evap} \sim 400$~GeV for the smallest objects with spherical shape that can attain hydrostatic equilibrium, $M \sim 10^{-10}~M_\odot$, to $m_{\rm evap} \simeq 0.7$~GeV for super-Jupiters and small brown dwarfs, $M \sim 10^{-2}~M_\odot$. For more massive brown dwarfs, the DM evaporation mass is slightly larger and grows when entering the stellar regime. Note, however, that for $M \gtrsim M_\oplus$, the DM evaporation mass only varies within an order of magnitude, $m_{\rm evap} \sim (1 - 10)$~GeV, and grows for smaller objects (rocky planets and satellites) due to their small size. This behavior follows the scaling $m_{\rm evap} \propto \tx R/(M \hat{\phi_c})$, with $\hat{\phi}_c \equiv v_e^2(r=0)/v_e^2(r=R)$, which can be understood from the fact that $E_c/\tx \sim 30$, as discussed in Section~\ref{sec:basics}. To illustrate the robustness of this result, we show the range $E_c/T_c = (20-40)$ with a band, which fully embeds the values of the DM evaporation mass for all objects and roughly accounts for systematics in modeling of celestial bodies properties. Furthermore, the small variation of the DM evaporation mass can be understood by considering the virial theorem, which implies that the factor $T_c \, R/M$ varies little for a given class of objects. Similarly to the jump in the equilibration time at $M = 2~M_\oplus$, the discontinuity at that value on the DM evaporation mass is caused by the abrupt transition in the composition and density profile, from rocky planets to icy planets and to the fact that $v_e < v_d$. For the considered parameters, equilibration is not reached for $M \lesssim 3 \times 10^{-8}~M_\odot$ (see Fig.~\ref{fig:taueq}) and this explains the slight bending of the curve towards smaller DM evaporation masses, as in those cases, the DM evaporation mass grows with time until reaching equilibrium. The same occurs for $M \gtrsim 60 \, M_\odot$, but in those cases equilibrium cannot be reached, as it would require a time longer than the age of those stars. Super-Earths, $M \lesssim 10~M_\oplus$, with a larger fraction of metals than what is assumed here, would have a slightly larger DM evaporation mass. Note, however, that for a given mass, heavier compositions generically imply smaller sizes~\cite{Fortney:2006jm, Seager:2007ix}.

Remarkably, as evident from Eqs.~(\ref{eq:eqmevapstars}) and~(\ref{eq:eqmevapplanets}), changes in parameters as cross sections, DM density or velocity dispersion, affect the DM evaporation mass only logarithmically. Therefore, given that $E_c/\tx \sim 30$, in order to obtain a value of the DM evaporation mass smaller by a factor of two, the term $\log \left(\rho_\chi \langle \sigma_A v_{\chi \chi} \rangle/v_d\right)$ for $v_e \gg v_d$, or $\log \left(\rho_\chi \langle \sigma_A v_{\chi \chi} \rangle/v_d^3\right)$ for $v_e \ll v_d$, must be larger by a factor of the order of $\sim (12 - 15)$. This implies that the sensitivity of the DM evaporation mass to changes on these parameters is relatively weak and thus, its value is rather stable against different particle physics models (with constant scattering cross sections) or for different locations of celestial bodies within the host galactic halo.

\begin{figure}[t]
	\centering
	\hspace*{-1.4cm}	\includegraphics[width=1.15\linewidth]{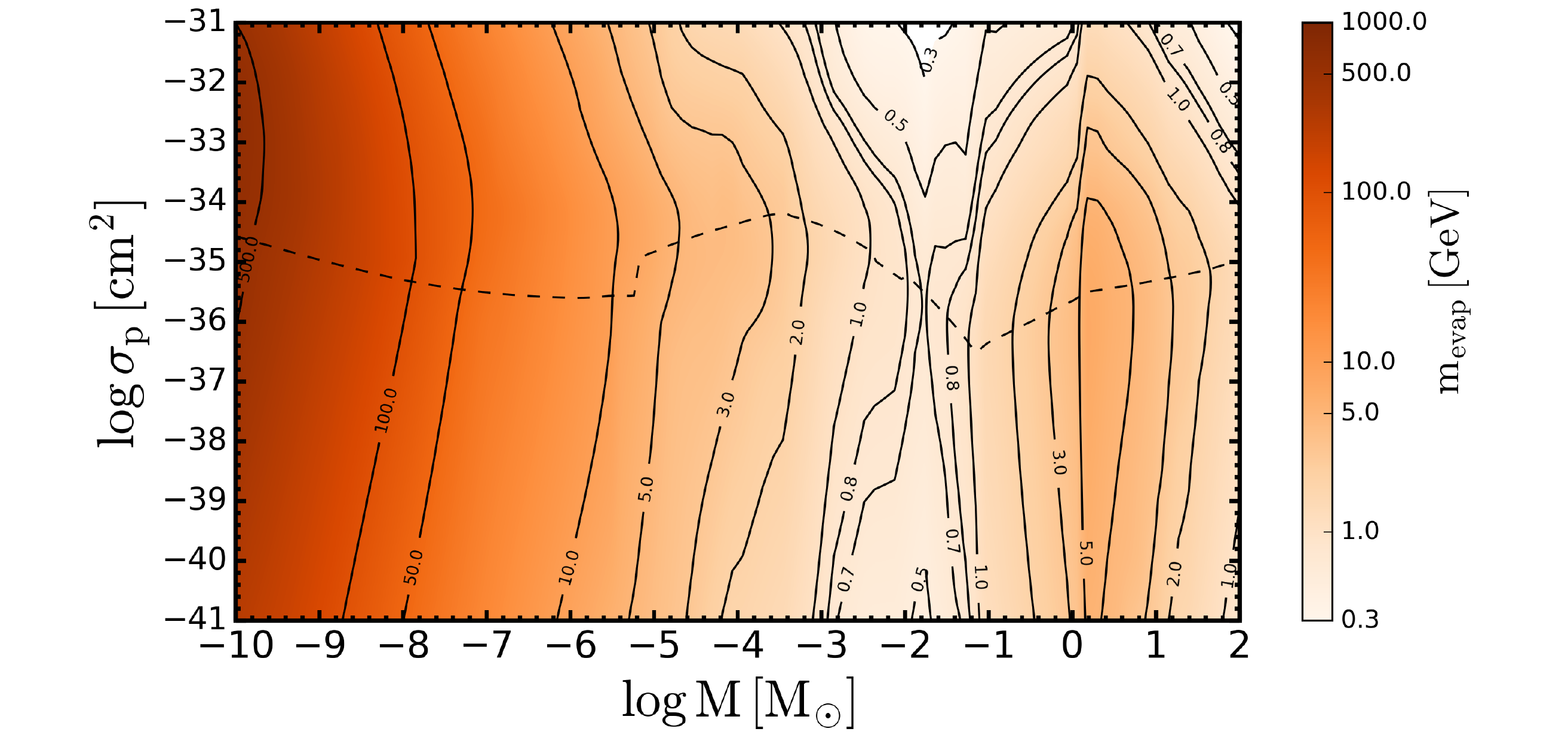}
	\caption{\textbf{\textit{2-D contour of the DM evaporation mass}} as a function of the SI scattering cross section and the mass of the capturing objects, for planetary bodies, brown dwarfs and main-sequence stars. We take the canonical value of the DM annihilation cross section $\langle \sigma_A v_{\chi \chi} \rangle = 3 \times 10^{-26}~\textrm{cm}^3/\textrm{s}$ and assume a position within the local neighborhood, $\rho_\chi = 0.4~\textrm{GeV}/\textrm{cm}^3$ and $v_d = 270~\textrm{km/s}$, although with other values similar results are obtained. The value of the SI geometric scattering cross section (dashed line), $\sum_i N_i \, \sigma_i^{\rm geom} = \pi \, R^2$, is approximately proportional to $R^2/M$, as expected.}  	
	\label{fig:evap2D}
\end{figure}

Nevertheless, the results in Fig.~\ref{fig:evap_geom} correspond to approximately the largest possible value of the DM evaporation mass for each celestial body (aside from variations on the average properties here considered), that is, the one obtained for a capture rate close to the saturation value. Next, we also study its variation with the value of the SI scattering cross section. This is depicted in Fig.~\ref{fig:evap2D}. The maximum value for all celestial bodies is visible in the figure, which is indeed close to that obtained for the geometric cross section (dashed line). For $M \gtrsim 2 \times 10^{-7}~M_\odot$, due to the exponential suppression of the evaporation rate in the thick regime, the smallest DM evaporation mass is achieved for the largest value of the scattering cross section we consider,\footnote{For larger cross sections, the assumed scaling with the atomic mass is not entirely reliable for contact interactions~\cite{Digman:2019wdm}.} $\sigma_p = 10^{-31} \, \textrm{cm}^2$. For $M \lesssim 2 \times 10^{-7}~M_\odot$, the transition between the thin and thick regimes takes place at values of the cross section closer to $\sigma_p = 10^{-31} \, \textrm{cm}^2$, so the smallest DM evaporation mass is found in the thin regime, for the smallest cross section we consider, $\sigma_p = 10^{-41} \, \textrm{cm}^2$. Note, however, that, for the average properties of celestial bodies considered in this work, the DM evaporation mass never reaches values below $\sim 250$~MeV in the parameter space shown in Fig.~\ref{fig:evap2D}. Therefore, even in extreme situations with $\rho_\chi \langle \sigma_A v_{\chi \chi} \rangle/v_d$ for $v_e \gg v_d$ (or $\rho_\chi \langle \sigma_A v_{\chi \chi} \rangle/v_d^3$ for $v_e \ll v_d$) being many orders of magnitude larger than what we have assumed for this figure, DM evaporation masses much below that mass are very unlikely. 

Along with the most massive stars, the objects for which the smallest values are obtained are super-Jupiters and small brown dwarfs, as correctly pointed out in Refs.~\cite{Leane:2020wob, Leane:2021ihh, Leane:2021tjj}. Nevertheless, in those papers, the estimated values of the DM evaporation mass down to a few MeV were obtained neglecting the critical exponential tail of the evaporation rate and thus, are incorrect and the conclusions reached for those very low masses are not valid; under the assumptions in this paper, the DM evaporation mass is rather $m_{\rm evap} > 250$~MeV in the entire parameter space shown in Fig.~\ref{fig:evap2D}. This can be understood from two facts: the critical exponential tail of the DM evaporation rate was incorrectly neglected, as explained in Section~\ref{sec:basics}, and the core temperature of brown dwarfs was underestimated by a factor $\gtrsim 4$ (cf. Refs.~\cite{Burrows:1997ka, Chabrier:2000iu, Burrows:2001rb, Becker:2014, Paxton:2010}). The importance of this tail, for the case of the Sun, is not a new finding, but has been known for over three decades~\cite{Gaisser:1986ha, Griest:1986yu, Gould:1987ju, Gould:1990}. We do stress that the fact that the DM evaporation mass is approximately given by $E_c/\tx \sim 30$ can be generalized to all spherical celestial bodies in hydrostatic equilibrium. Indeed, note that Ref.~\cite{Zentner:2011wx} did correctly estimate the DM evaporation mass for the most massive brown dwarfs and low-mass stars in the context of asymmetric DM scenarios. Additionally, notice that we have considered temperatures of brown dwarfs corresponding to late evolutionary stages. From this point of view, our assumption is conservative, as younger brown dwarfs are warmer, which results in a higher DM evaporation mass. Note that even older ($\gtrsim 10$~Gyr) brown dwarfs would have a slightly lower core temperature ($\lesssim 20\%$ cooler than at $5$~Gyr for the most massive ones and even more similar for the least massive ones), which implies a correspondingly lower DM evaporation mass. On another hand, let us stress again that uncertainties on the density or temperature profiles have a small impact on the DM evaporation mass. All in all, these uncertainties are approximately accounted for by the gray band in Fig.~\ref{fig:evap_geom}.

As we have discussed above and can be seen from Fig.~\ref{fig:evap2D}, the DM evaporation mass for a given object mass is rather stable, within about a factor of two at most, against variations by ten orders of magnitude in the scattering cross section. Therefore, this is a robust result. We emphasize, however, that these results correspond to simplified and average properties of all the celestial bodies we consider. Given that, for the geometric value of the scattering cross section, $m_{\rm evap} \simeq 30 \, \tx R/(G \, M \, \hat{\phi}_c)$, the uncertainty in the DM evaporation mass is also driven by the scatter over the properties of the capturing objects. As mentioned above, the virial theorem implies a small variation of $T_c \, R/M$ for a given class of objects, and thus, a small variation of the DM evaporation mass.

\subsection{Further comments on the dependence of the DM evaporation mass on cross sections}

All the above results are obtained for SI scattering cross sections, such that DM particles couple to the nuclei mass. In the case of SD interactions, DM couples to the spin of the target, so not all nuclei could contribute to the capture and evaporation processes. Moreover, SD interactions are not enhanced by the coherence factor $A_i^2$, as happens in the SI case. This is particularly important for planetary bodies, made up mainly of silicates and metals. For solar abundances, only a small fraction of their elements, $\lesssim 1\%$, could contribute to DM scattering via SD interactions. This definitely results in equilibration times which are longer than for SI interactions by a few orders of magnitude for similar DM-nucleon cross sections, which can prevent the system to reach this state. Nevertheless, the logarithmic dependence of the DM evaporation mass on $\sum_i N_i \, \sigma_i$ implies that it is only reduced by $\lesssim 20\%$ even for these objects. For celestial bodies with large amounts of hydrogen, such as giant planets, brown dwarfs and stars, the differences between SI and SD interactions are even smaller (see, e.g., the differences for the case of the Sun~\cite{Garani:2017jcj}). 

Also note that throughout this work we have not considered the effect of self-scatterings~\cite{Zentner:2009is}. The presence of additional interactions among DM particles would enhance the capture rate by introducing a source of self-capture~\cite{Zentner:2009is}, but it would also enhance the evaporation rate by introducing a source of self-ejection~\cite{Gaidau:2018yws} and of self-evaporation~\cite{Chen:2014oaa}, the three processes being proportional to the same self-interaction cross section. As a result, in the thin regime, the DM evaporation mass is slightly larger than in the usual case without self-interactions~\cite{Chen:2014oaa, Gaidau:2018yws}. This trend can be qualitatively understood from the dependence of the DM evaporation mass on the scattering cross section in the absence of self-interactions. Larger cross sections imply larger DM evaporation masses. In the thick regime, the dependence on the cross section is reversed and larger cross sections result in lower DM evaporation masses, as can be seen from Fig.~\ref{fig:evap2D}. Nevertheless, in the thick regime, the effect of self-interactions is unlikely to have a significant impact because for that to occur,  these interactions would have to be stronger than current bounds~\cite{Zentner:2009is}. Therefore, generically, the presence of self-interactions would tend to increase the value of the DM evaporation mass.

Finally, small DM evaporation masses have been claimed~\cite{Leane:2020wob} within the Co-SIMP scenario~\cite{Smirnov:2020zwf}, in which the DM freeze-out is assisted by Standard Model (SM) particles, $\chi + \chi + {\rm SM} \to \chi + {\rm SM}$. The annihilation rate in this case is defined as
\begin{equation}
\mathcal{A}_{\rm Co-SIMP} =  \frac{\int_0^{R} n_\chi^2(r,t) \, n_{\rm SM} \, \langle \sigma_{3 \to 2} \, v_{\chi \chi}^2\rangle \, 4 \pi \, r^2
	\, \dd r }{\left(\int_0^{R} \, n_\chi(r,t) \, 4\pi \, r^2 \, \dd r \right)^2}  \simeq \frac{\langle \sigma_{3 \to 2} \, v_{\chi \chi}^2\rangle \, n_{\rm SM}}{V_s} \simeq \frac{\langle \sigma_{3 \to 2} \, v_{\chi \chi}^2\rangle}{V_s^2} \sum_i  \frac{0.1 \, M}{m_i} ~,
\end{equation}
where $n_{\rm SM} \simeq \sum_i N_i/V_s$ is the number density of SM particles participating in the process. For the typical case of the Co-SIMP scenario, $\mx \ll m_i$, the cross section of the number-changing interaction required to obtain the observed value of the relic density is $\langle \sigma_{3 \to 2} \, v_{\chi \chi}^2\rangle \simeq 10^3 \, ({\rm GeV}/\mx)^3 \, {\rm GeV}^{-5}$~\cite{Smirnov:2020zwf}. To compare the annihilation rate of this process to that of standard $2 \to 2$ annihilations, we take $m_i = m_p$ and estimate
\begin{equation}
\langle \sigma_{3 \to 2} \, v_{\chi \chi}^2\rangle  \, n_{\rm SM} \simeq 8 \times 10^{-30} \, {\rm cm}^3/{\rm s} \, \left(\frac{\langle \sigma_{3 \to 2} \, v_{\chi \chi}^2\rangle}{10^3 \, ({\rm GeV}/\mx)^3 \, {\rm GeV}^{-5}}\right) \, \left(\frac{M}{M_\odot}\right) \, \left(\frac{R_\odot}{R}\right)^3 \, \left(\frac{\rm GeV}{\mx}\right)^3 ~.
\end{equation}
Therefore, for $\mx \gtrsim 100$~MeV, the annihilation rate is smaller than for the usual $2 \to 2$ scenario, with the canonical value of the annihilation cross section we have used throughout this work, $\langle \sigma_{\rm A} \, v_{\chi \chi}\rangle = 3 \times 10^{-26}~{\rm cm}^3/{\rm s}$. Thus, within the Co-SIMP scenario, the equilibration time is longer. As a consequence, when equilibrium is reached, the DM evaporation mass within the Co-SIMP scenario is slightly larger than for the usual $2 \to 2$ annihilations, in contrast to recent claims~\cite{Leane:2020wob}.

\subsection{DM evaporation mass in post-main-sequence stars and compact objects}

As the internal properties of stars after they leave main sequence change significantly in short periods of time, the calculation of the minimum mass of DM particles that can be efficiently trapped during these stages becomes non-trivial and highly time dependent. The study of this time dependence is beyond the scope of this work, but the general trend is that the DM evaporation mass grows in time with respect to its value at main sequence during stages with an inert core and it is similar during periods with a burning core. At the last stage of the life of stars, when nuclear fusion cannot take place any more and they live on as cool compact remnants, with a high escape velocity, the DM evaporation mass is significantly reduced.

We do not attempt a full study of the post-main-sequence evolution of the DM evaporation mass, but we qualitatively describe it based on the simplified description presented in the previous section. Whenever capture and annihilation processes are not in equilibrium, the DM evaporation mass would grow, attaining its maximum value at equilibrium, if the properties of the star do not significantly change and if the DM thermalization time is short enough. Note that during main sequence, the equilibration time is $\tau_{\rm eq}\sim (10^5 - 10^6)$~yr, for the parameters considered in Fig.~\ref{fig:taueq}, so in general, equilibrium could be reached during the post-main-sequence phases. Therefore, the discussion can be driven by the robust result we found for all spherical objects in hydrostatic equilibrium: the DM evaporation mass is approximately determined by $E_c/\tx \sim 30$, which implies that it scales as $m_{\rm evap} \propto \tx R/(M \hat{\phi_c})$. Recall that the average DM temperature inside celestial bodies is very close to their core temperature.

As a general trend, whenever a star has an active shell burning material surrounding an approximately inert core, core contraction results in envelope expansion and core expansion in envelope contraction. The former implies larger radius and $\hat{\phi}_c$ and a hotter core, whereas the latter implies the opposite.

\begin{figure}[t]
	\centering
	\hspace*{-0.65cm}	\includegraphics[width=1.05\linewidth]{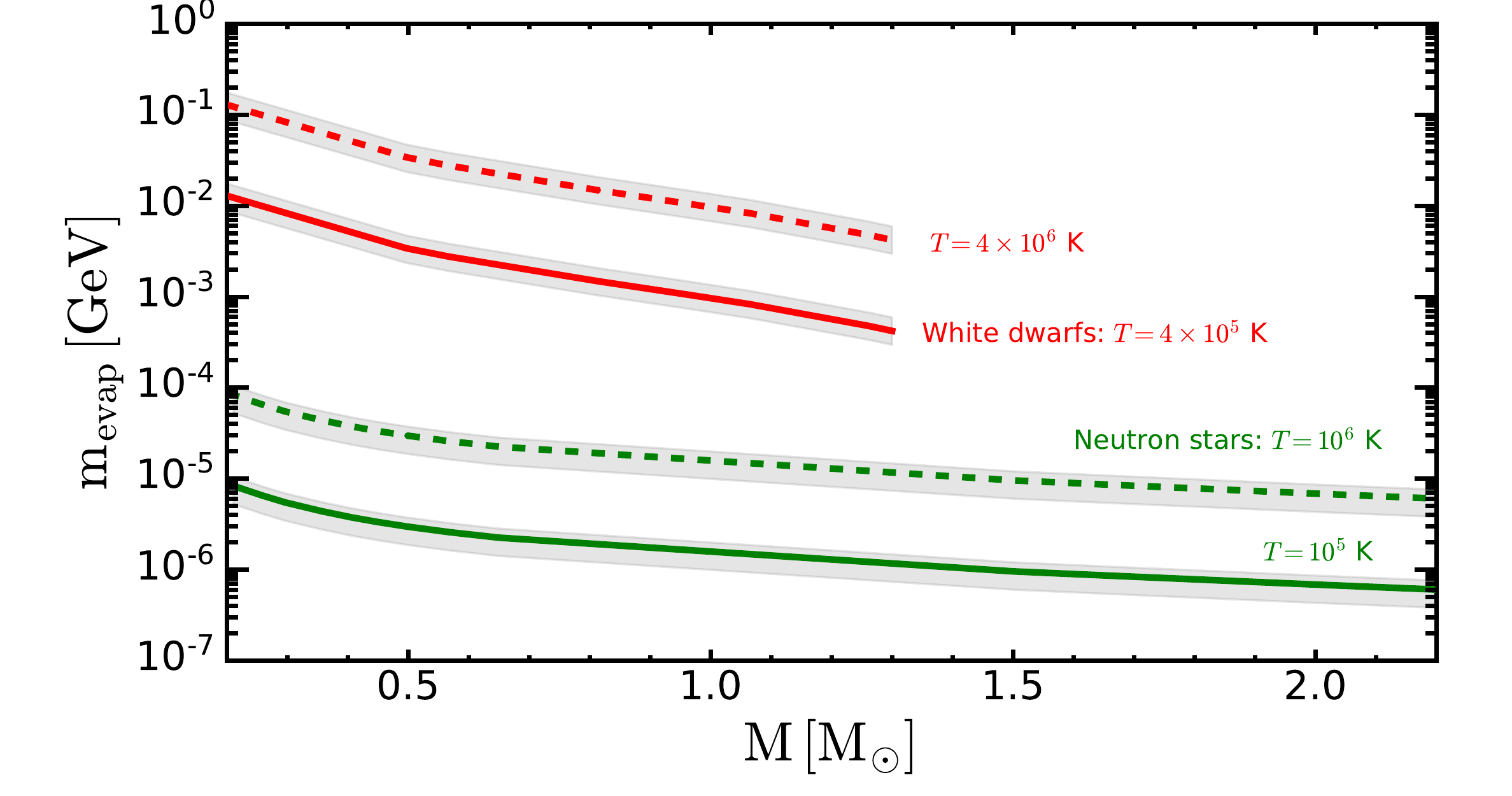}
	\caption{\textbf{\textit{DM evaporation mass for compact objects}}, white dwarfs (red upper lines) and neutron stars (green lower lines), for two  temperatures: $T = 4 \times 10^5$~K (red solid line) and $T = 4 \times 10^6$~K (red dashed line) for white dwarfs; and $T = 10^5$~K (green solid line) and $T = 10^6$~K (green dashed line) for neutron stars. We take the geometric cross section, $\sum_i N_i \, \sigma_i^{\rm geom} = \pi \, R^2$, and $\langle \sigma_A v_{\chi \chi} \rangle = 3 \times 10^{-26}~\textrm{cm}^3/\textrm{s}$ for white dwarfs and $\langle \sigma_A v_{\chi \chi} \rangle = 0$ for neutron stars, and assume a position within the local neighborhood, $\rho_\chi = 0.4~\textrm{GeV}/\textrm{cm}^3$ and $v_d = 270~\textrm{km/s}$. The shaded bands depict the range $E_c/T_c = (20 - 40)$.}  	
	\label{fig:DMevapcompact}
\end{figure}

For low-mass stars, $M \lesssim 2.3~M_\odot$, during their phase as subgiants, the core temperature and radius are slightly larger than those during main sequence, and their mass is similar. Given the larger density gradient, the ratio of escape velocities at core and surface is slightly larger than during main sequence. This implies that the DM evaporation mass during this phase does not significantly change. While climbing the red giant phase, the core temperature increases up to $\sim 10^8$~K, the mass slightly decreases, the radius increases by up to two orders of magnitude and the core density by several orders of magnitude, increasing $\hat{\phi}_c$. Assuming DM thermalization to be fast enough, this results in a decrease of the DM evaporation mass with respect to main sequence, which grows in time, though. Moving to the horizontal branch, material is burnt not only in a shell surrounding the core, but in the core itself. Moreover, the star contracts, reducing $\hat{\phi}_c$. The higher DM temperature and the smaller $\hat{\phi}_c$ with respect to main sequence implies a slightly larger DM evaporation mass in the thin regime, although it is similar in the thick regime~\cite{Gould:1990}. During the asymptotic giant branch, the DM evaporation mass follows a similar trend to that during the red giant branch, i.e., it tends to decrease first and then grows in time. From the tip of the asymptotic giant branch on, stars lose mass (planetary nebulae), contract and cool. The resulting effect would be a reduction of the DM evaporation mass. Note, however, that this phase lasts too short for capture and annihilation to reach equilibrium (or even for DM to thermalize), so the time scale for the decrease of the DM evaporation mass is too long to be effective. As a consequence, there is an abrupt reduction in the DM evaporation mass from the end of the asymptotic giant branch to the white dwarf phase, with values  $m_{\rm evap} \lesssim 130$~MeV for white dwarfs older than $\sim 3$~Gyr (or $T_c \lesssim 4 \times 10^6$~K).  Also note that as they cool down as white dwarfs, the DM evaporation mass continues decreasing. This is illustrated in Fig.~\ref{fig:DMevapcompact}, where we show the DM evaporation mass for white dwarfs with core temperatures $T_c = 4 \times 10^5$~K and $T_c = 4 \times 10^6$~K.

More massive stars spend much less time in post-main-sequence phases, and in general, DM capture and annihilation would not reach equilibrium and even DM might not completely thermalize. Their non-degenerate helium core keeps on burning without going through the helium flash. Intermediate-mass stars, $2.3~M_\odot \lesssim M \lesssim (8-10)~M_\odot$, while burning helium, cross the instability strip and experience significant structural changes, before reaching the asymptotic giant branch. Similarly to stars in the horizontal branch, the DM evaporation mass in these intermediate-mass stars would be similar to that of their main-sequence parents (larger in the thin regime due also to the heavier composition) and then it would tend to decrease when getting to the asymptotic giant branch, although again, the time spent in this phase might be too short. The endpoint of this evolution is the most massive white dwarf stage, $M \gtrsim 0.6~M_\odot$, with $m_{\rm evap} \lesssim 30$~MeV (when older than $\sim 3$~Gyr, or $T_c \lesssim 4 \times 10^6$~K).

As can be seen from Fig.~\ref{fig:DMevapcompact}, the DM evaporation mass for white dwarfs varies from $\sim 130$~MeV to $\sim 0.4$~MeV (for the geometric scattering cross section and the canonical annihilation cross section) within the core temperature range we consider.\footnote{After the first version of our paper appeared on the arXiv, another calculation of the DM evaporation mass for white dwarfs was presented~\cite{Bell:2021fye}. When correctly accounting for the differences (mainly the core temperature, but also the density profile and the fact that their calculation is equivalent to that for asymmetric DM scenarios), those results are in agreement with ours. Nevertheless, note that simply correcting for the core temperature results in similar DM evaporation masses, within a factor of two.} This variation is about two orders of magnitude larger, within the small mass range for white dwarfs, $0.2~M_\odot \lesssim M \lesssim 1.3~M_\odot$, than that for main-sequence stars, within a much larger mass range, $0.07~M_\odot \lesssim M \lesssim 100~M_\odot$. In addition to the effect of cooling, this can be understood from the mass--radius relation of white dwarfs, depicted in the left panel of Fig.~\ref{fig:MRR-compact}. 

The most massive stars, $M \gtrsim (8 - 10)~M_\odot$, keep on burning heavier elements while growing in size, maintaining their luminosity rather unchanged. Nevertheless, they spend a very short period of time until they cannot burn material anymore, so equilibration between DM capture and annihilation is not reached in general. In any case, given that $E_c/\tx$ only varies within a factor of a few during the burning phase, the DM evaporation mass is not expected to change much before these stars end up as neutron stars (or black holes).\footnote{An example of the evolution of the DM capture and annihilation processes in massive stars is provided in Ref.~\cite{Brdar:2016ifs}.} After the core collapses into a neutron star (if it is a black hole, the concept of DM evaporation mass is meaningless), there is an abrupt decrease of the DM evaporation mass, reaching values as low as $m_{\rm evap} \simeq 0.6$~keV for the heaviest and coolest neutron stars and $m_{\rm evap} \simeq 80$~keV for the least massive and hottest neutron stars. This is illustrated in Fig.~\ref{fig:DMevapcompact}, where the variation of the DM mass is clearly less pronounced than for white dwarfs. Likewise, this can be understood from the mass--radius relation of neutron stars, depicted in the right panel of Fig.~\ref{fig:MRR-compact}. As already discussed in Section~\ref{sec:basics}, the DM annihilation cross section must be very small for DM particles with keV mass to not overclose the Universe. Therefore, we compute the DM evaporation mass in the limit $\tau_{\rm eq} \to \infty$, equivalent to asymmetric DM scenarios~\cite{Garani:2018kkd}. Note that, in general, the DM evaporation mass in asymmetric scenarios is larger than in symmetric ones when equilibrium is reached.

\section{Summary and conclusions}
\label{sec:conclusions}

The effects of capture of DM particles by celestial bodies have been extensively studied in the literature during the last decades. Even if DM particles scatter with the medium and are finally gravitationally trapped within an object, it turns out that light DM particles are very likely to be quickly kicked out and escape. This is the process of DM evaporation, which sets a minimum DM mass that could guarantee a stable population of trapped DM particles. In this work, we have computed in detail the DM evaporation mass for all spherical celestial bodies in hydrostatic equilibrium, assuming constant scattering cross sections. For planetary bodies, brown dwarfs and main-sequence stars, spanning the mass range $10^{-10}~M_\odot \leq M \leq 10^2~M_\odot$, we obtain the DM evaporation mass for a wide range of DM-nucleon SI cross sections, $10^{-41}~\textrm{cm}^2 \leq \sigma_p \leq 10^{-31}~\textrm{cm}^2$. For the average properties of celestial bodies we consider, at the local galactic position, the absolute minimum for the DM evaporation mass is $m_{\rm evap} \simeq 250$~MeV, for the most massive stars and the largest cross sections. For super-Jupiters and low-mass brown dwarfs, a minimum value of $m_{\rm evap} \simeq 300$~MeV is obtained for the largest cross section we consider. For very compact objects, such as white dwarfs and neutron stars, smaller DM evaporation masses are found, with values as low as $m_{\rm evap} \simeq 0.4$~MeV (for $T_c = 4 \times 10^5$~K) and $m_{\rm evap} \simeq 0.6$~keV (for $T_c = 10^5$~K), respectively. These limiting values for the DM evaporation mass correspond to the canonical value of the ($s$--wave) annihilation cross section and at our local galactic position, although the dependence on these parameters is only logarithmic.

In Section~\ref{sec:basics}, we have defined the concept of DM evaporation mass and have introduced all the required ingredients for its calculation. We have discussed the critical importance of the exponential tail of the DM evaporation rate (Fig.~\ref{fig:tails}), which had already been studied for the case of the Sun~\cite{Gaisser:1986ha, Griest:1986yu, Gould:1987ju, Gould:1990}, although its importance has not always been appreciated. These early papers obtained a DM evaporation mass for the Sun which is approximately given by $E_c/\tx \simeq 30$, where $E_c$ is the escape energy at the core of captured DM particles and $\tx$ is their temperature. Similar values are found for the DM evaporation masses obtained for the Earth~\cite{Freese:1985qw, Krauss:1985aaa, Gould:1988eq, Garani:2019rcb} and the Moon~\cite{Garani:2019rcb}. This estimate corresponds to the geometric cross section, $\sum_i N_i \, \sigma_i^{\rm geom} = \pi \, R^2$. Here, we generalize this result for all round celestial bodies in hydrostatic equilibrium. The virial theorem is at the core of this finding.

In Section~\ref{sec:celestialbodies} we have described the average properties, relevant for the calculation of the DM evaporation mass, of all celestial bodies we consider throughout the paper: planetary bodies, brown dwarfs, main-sequence stars, post-main-sequence phases of stellar evolution, white dwarfs and neutron stars. We have provided mass--radius (Figs.~\ref{fig:MRR} and~\ref{fig:MRR-compact}) and mass--core temperature (Fig.~\ref{fig:MTR}) relations, as well as density and temperature profiles and composition, to describe the average properties of all these objects. The derived escape velocity at the surface, along with two reference values for the galactic DM dispersion velocity, is depicted in Fig.~\ref{fig:Mve}. 

Finally, in Section~\ref{sec:results} we have discussed the DM equilibration time (Fig.~\ref{fig:taueq}) and have computed the DM evaporation mass for all these celestial bodies, as a function of the mass of the object (Figs.~\ref{fig:evap_geom} and~\ref{fig:DMevapcompact}) and of the SI scattering cross section (Fig.~\ref{fig:evap2D}). We have also discussed the dependence with other parameters, as the position of the celestial body in the galactic halo (DM density and velocity), the DM annihilation cross section, and the type of interaction (SI and SD). The DM evaporation mass, however, depends only logarithmically on these parameters, so its value is rather stable against variations of them. We have also commented on the impact of DM self-interactions or non-canonical DM annihilation processes on the DM evaporation mass.

For the geometric value of the scattering cross section, the minimum value of the DM evaporation mass is obtained for super-Jupiters and low-mass brown dwarfs (Fig.~\ref{fig:evap_geom}), $m_{\rm evap} \simeq 0.7$~GeV (at our local galactic position). The fact that these objects are optimal sites to search for effects of capture of light DM particles has been pointed out recently~\cite{Leane:2020wob, Leane:2021ihh, Leane:2021tjj}, under the assumption of constant scattering cross sections. Nevertheless, those papers neglected the crucial exponential tail of the evaporation rate and estimated a DM evaporation mass as low as $\sim 4.5$~MeV, which represents an underestimation of the correct result by at least one order of magnitude, even after accounting for uncertainties on the modeling of celestial bodies. Similarly, a too low DM evaporation mass for planets has also been suggested using similar arguments~\cite{Bramante:2019fhi}. Therefore, we argue that the conclusions reached in those papers for masses below the correctly evaluated (properly accounting for the exponential tail) DM evaporation mass are not valid. 

Finally, we stress again the general and robust result we obtain: for constant scattering cross section at the geometric value, at our local galactic position, the DM evaporation mass for all spherical celestial bodies in hydrostatic equilibrium is approximately given by the simple expression $E_c/\tx \sim 30$, which provides the correct result within $\lesssim 30\%$ in the mass range $10^{-10}~M_\odot \leq M \leq 10^2~M_\odot$ and in the SI scattering cross section range \linebreak $10^{-41}~\textrm{cm}^2 \leq \sigma_p \leq 10^{-31}~\textrm{cm}^2$. The dependence on the local galactic DM density, velocity, and on the scattering and annihilation cross sections is only logarithmic, and uncertainties on the interior density and temperature profiles of celestial bodies have a small impact.

\section*{Acknowledgments}
RG is supported by MIUR grant PRIN 2017FMJFMW. SPR is supported by the Spanish FEDER/MCIU-AEI grant FPA2017-84543-P and  MCIN/AEI/10.13039/501100011033 grant PID2020-113334GB-I00, and partially, by the Portuguese FCT (UID/FIS/00777/2019 and CERN/FIS-PAR/0004/2019). SPR also acknowledges support from the European ITN project HIDDeN (H2020-MSCA-ITN-2019//860881-HIDDeN). The authors thank the Galileo Galilei Institute for hospitality.

\bibliography{biblio_evapmass}

\end{document}